\documentclass[traditabstract]{aa} 

\newcommand{\unit}[1]{\ensuremath{\mathrm{\thinspace #1}}}
\newcommand{\od}[2]{\frac{\mathrm{d}#1}{\mathrm{d}#2}}
\newcommand{\reffig}[1]{\figurename~\ref{#1}}
\newcommand{\reftab}[1]{\tablename~\ref{#1}}

\usepackage{slashbox} 

\def\mearth{\mathrm{\thinspace M_\oplus}}
\def\rearth{\mathrm{\thinspace R_\oplus}}

\newcommand{\vect}[1]{\overrightarrow{#1}}

\usepackage{graphicx}
\usepackage{subfig}
\usepackage{txfonts}

\usepackage{natbib}
\usepackage{aas_macros}

\newcommand{\modif}[1]{#1}

\begin{document}

\titlerunning{Super-Earths and giant planet cores}
\authorrunning{Cossou et al.}

\title{Hot super-Earths and giant planet cores from different migration histories}

\author{Christophe Cossou\inst{1,2}, Sean N. Raymond\inst{1,2}, Franck Hersant\inst{1,2} \and Arnaud Pierens\inst{1,2}}

\institute{
Univ. Bordeaux, Laboratoire d'Astrophysique de Bordeaux, UMR 5804, F-33270, Floirac, France.
\and
CNRS, Laboratoire d'Astrophysique de Bordeaux, UMR 5804, F-33270, Floirac, France.\\
\email{rayray.sean@gmail.com}}

\date{21 Apr 2014 / 19 July 2014}

\abstract{Planetary embryos embedded in gaseous protoplanetary disks undergo Type I orbital migration. Migration can be inward or outward depending on the local disk properties but, in general, only planets more massive than several $M_\oplus$ can migrate outward.  Here we propose that an embryo's migration history determines whether it becomes a hot super-Earth or the core of a giant planet. Systems of hot super-Earths (or mini-Neptunes) form when embryos migrate inward and pile up at the inner edge of the disk.  Giant planet cores form when inward-migrating embryos become massive enough to switch direction and migrate outward. We present simulations of this process using a modified N-body code, starting from a swarm of planetary embryos.  Systems of hot super-Earths form in resonant chains with the innermost planet at or interior to the disk inner edge.  Resonant chains are disrupted by late dynamical instabilities triggered by the dispersal of the gaseous disk. Giant planet cores migrate outward toward zero-torque zones, which move inward and eventually disappear as the disk disperses. Giant planet cores migrate inward with these zones and are stranded at $\sim 1-5$~AU.  Our model reproduces several properties of the observed extra-solar planet populations.  The frequency of giant planet cores increases strongly when the mass in solids is increased, consistent with the observed giant exoplanet - stellar metallicity correlation.  The frequency of hot super-Earths is not a function of stellar metallicity, also in agreement with observations.  Our simulations can reproduce the broad characteristics of the observed super-Earth population.}


\maketitle

\section{Introduction}

One third to one half of all main sequence FGKM stars host planets slightly larger or more massive than Earth on orbits with periods shorter than 50-100 days~\citep{mayor11,howard10,howard12,fressin13,howard2013observed,dong13,petigura13}.  Planets smaller than $1.5-2 R_\oplus$ are likely to be rocky ``super-Earths'' whereas larger planets are more likely to have thick gaseous envelopes and be ``mini-Neptunes''~\citep{weiss13,lopez13,weiss2014mass,marcy2014masses,marcy2014occurence}.  Super-Earths and mini-Neptunes are usually found in systems with many planets~\citep[e.g.][]{mayor09,lovis11,lissauer2011closely}.  The orbital architecture of these systems is generally compact but non-resonant, and confined within a few tenths of an AU of their host stars~\citep{lissauer2011architecture,fabrycky2012transit}. 

On the other hand, gas giant planets close to their stars are rare.  Only 0.5-1\% of stars host a hot Jupiter~\citep[][]{cumming08,howard10,howard12,mayor11, wright2012frequency}.  There is a dearth of gas giants with orbital radii less than $\sim 0.5$~AU.  From 0.5 to 1 AU there is a rapid increase in gas giant frequency and a plateau out to 3-5 AU.  Radial velocity surveys show that at least 14\% of Sun-like stars host a gas giant with a period less than 3000 days~\citep{mayor11}.  Microlensing surveys hint that the abundance of long-period gas giants may be as high as 50\%~\citep{gould10}.

In the core-accretion model, gas giants form in two steps~\citep{mizuno1980progress,pollack96,ida04,rice2004accelerated,alibert2006formation,mordasini09}.  First a solid core of $5-10 \mearth$ forms.  Then the core gravitationally captures gas from the disk.  Capture of gas is initially slow and limited by the core's ability to cool and contract~\citep{ikoma2000formation,hubickyj2005accretion}.  Once the envelope mass becomes comparable to the core mass gas accretion can enter a runaway phase and the planet can grow to Jupiter-size within $\sim 10^5$ years~\citep{lissauer2009models}.

Uranus and Neptune are often considered to be giant planet cores that failed to undergo runaway gas accretion~\citep[e.g.,][]{helled13}.  Given their significant gas contents, the ice giants' accretion must have taken place during the gaseous disk phase.  By the same argument, mini-Neptunes (with $R > 1.5-2 \rearth$) may also be considered failed giant planet cores.  Given that hot super-Earths and mini-Neptunes form a reasonably continuous distribution, it stands to reason that hot super-Earths represent lower-mass actors in the same story.  

Orbital migration must play a key role in shaping the destinies of growing planetary embryos.  Mars-mass or larger objects launch density waves in gaseous protoplanetary disks~\citep{goldreich1979excitation}.  The waves torque the planets' orbits and cause Type I orbital migration~\citep{goldreich80,ward86}.  Spiral density waves generated by Lindblad resonances generate a negative torque that tends to drive planets inward~\citep{ward97,tanaka02}.  However, density perturbations in the co-orbital region generate a positive torque that can under some circumstances counteract the differential Lindblad torque~\citep{paardekooper06,kley08,paardekooper2010torque,paardekooper2011torque}.  Outward migration only occurs for planets more massive than $\sim 5 \mearth$, although this critical mass is a strong function of the disk properties~\citep{bitsch13,bitsch2014stellar}.  The region of outward migration does not extend all the way to the star (see \reffig{fig:migration_map} below); within $\sim 1$~AU migration is almost universally directed inward. \modif{Narrow regions of outward migration can exist closer-in (e.g., from 0.3-0.5 AU).}

We propose that low-mass planets (both hot super-Earths and mini-Neptunes) and the cores of gas giant planets originate from the same parent population of primordial planetary embryos.  The bifurcation between the two populations is the result of a competition between growth and orbital migration.  In the limit of very slow growth, planetary embryos migrate inward.  Rather than collide with the star, their migration is halted by a corotation torque generated by strong temperature and density gradients at the inner edge of the disk~\citep{masset06}.  However, if an embryo grows massive enough at a large enough orbital radius then it can transition from inward to outward migration.  It then settles in a zero-torque zone at several AU~\citep{lyra10}.  Given that the zero-torque radius evolves slowly, the core has time -- and a steady supply of food in the form of inward-migrating embryos -- to grow into a gas giant.  

The paper is structured as follows.  We first review current models for the formation of hot super-Earths and giant planet cores in Section~\ref{sec:review}.  In Section~\ref{sec:methods} we present our model and its implementation.  In Section~\ref{sec:static} we present the results of simulations in static disks that do not evolve.  These simulations serve to illustrate the key phenomena at play.  In Section~\ref{sec:dissip} we present the results of simulations in dissipating disks.  We compare the planetary systems generated by the simulations with observations.  We compare both the characteristics of observed hot super-Earths and the frequency of gas giant planets.  In Section~\ref{sec:discussion} we discuss the pros and cons of our model and conclude.  

\section{Review of formation models}\label{sec:review}

Planet formation models are in a state of continual evolution.  \modif{Observations and theory are two key pieces of the puzzle.}  Measurements provide new data-points, and new ideas are generated to match them.

\subsection{Accretion of hot super-Earths}
In recent years a plethora of models has been proposed to explain the origin of systems of hot super-Earths~\citep[see][]{raymond2008observable,raymond14}.  Several models required specific planetary system architectures that differed from those observed.  These models were discarded.  The two surviving candidate models are in-situ accretion and inward migration.  

The in-situ accretion model proposes that hot super-Earths grow in a bottom-up fashion from local material very close-in to their stars.  This model was proposed by \cite{raymond2008observable}, who discarded it as requiring unrealistically large disk masses.  The model was revived by several recent papers.  Indeed, \cite{hansen12,hansen13} showed that in-situ accretion can explain several aspects of the observed Kepler candidates.  This includes both the orbital spacing of adjacent planets and their expected eccentricity and inclination distributions.  \cite{chiang13} showed that the observed hot super-Earths can be used to build a ``minimum-mass extrasolar nebula'', a universal disk from which such planets would presumably have formed.  However, \cite{raymond2014universal} showed that multiple planet systems show no sign of a universal minimum-mass disk.  In fact, a significant fraction of inferred minimum-mass disks have slopes that are so shallow or steep that they are inconsistent with accepted disk physics.  In addition, \cite{chiang13}'s minimum-mass nebula is an order of magnitude larger than the one expected from the Solar System or from sub-mm observations of outer disks.  

The inward migration model was proposed by \cite{terquem2007migration} and built upon by \cite{ogihara09}, \cite{mcneil09} and \cite{cossou2014making}.  It proposes that hot super-Earths form at large orbital radii, type I migrate inward and grow by embryo-embryo collisions.  The model requires that roughly Earth-sized embryos grow while the gaseous disk is still present.  Of course, growth is already required to explain the formation of gas- and ice-giant planets.  The model generally produces systems containing many planets in long chains of mean motion resonances.  Resonances can be broken either by late instabilities~\citep{terquem2007migration}, turbulence~\citep{rein12} or simply long-term orbital dynamics~\citep{goldreich14}. The inward migration model is favored both to explain the inflated radii of mini-Neptunes~\citep{rogers11} and the orbital architecture of a number of systems~\citep[e.g.][]{rein12,swift13}.

In the inward migration model, the inner disk is enhanced in solids by objects large enough to undergo type I migration.  Of course, one can imagine a scenario in which the inner parts of protoplanetary disks were enriched in solids when objects were smaller.  This could be in the form of small particles~\citep{chatterjee2014inside,boley13} or even m-sized boulders~\citep{weidenschilling77} that undergo rapid inward radial drift.  While such a scenario is appealing, it remains to be adequately vetted.

In this paper we build upon the inward migration model.  Although the in-situ accretion model can match several aspects of the observations, the model is built on a flimsy foundation.  By assuming that the inner parts of protoplanetary disks are an order of magnitude more massive than expected and with strange density profiles, the in-situ accretion model is hard to accept.  As we show below, late instabilities in systems of migrated hot super-Earths can provide a transition to a similar regime as the in-situ model and reproduce the same observations.

\subsection{Growth of giant planet cores}

The growth of giant planet cores have not been successfully modeled.  \cite{thommes03} showed that standard, bottom-up oligarchic growth cannot form reasonable cores within a typical disk lifetime.  \cite{levison10} showed that large cores can in some cases grow rapidly when planetesimal-driven migration is taken into account.  However, large cores were only produced in 10\% of their simulations, even in very massive disks. The problem is that the Hill radius $R_H = a (M/3M_\star)^{(1/3)}$ scales more strongly with orbital radius than with planet mass. Therefore, as a planet grows its ability to scatter nearby planetesimals outstrips its ability to accrete them.  The growing embryo is thus a victim of its own initial success~\citep[see, for example, ][]{levison01}. \cite{lambrechts12} and \cite{morby12} showed that cores may grow efficiently by accreting cm-sized pebbles. Observations have shown that pebbles do indeed exist in protoplanetary disks \citep{wilner2005toward}. \modif{What remains to be determined is the size distribution of growing objects in a typical disk, i.e., whether cores are built from planetesimals or pebbles.} 

Given that embryos migrate inward in some parts of the disk and outward in other (see below), there exist zero-torque locations sometimes called ``convergence zones''.  It has been proposed that gas-driven migration can assemble planetary embryos at these locations~\citep{lyra10,horn12,hellary2012global}.  However, embryos rarely accrete in these zones.  Rather, they tend to form long resonant chains~\citep{morby08,sandor11,cossou2013convergence}.  Turbulence can destabilize resonant chains if they get long enough and produce $5-10 \mearth$ cores~\citep{horn12,pierens13}.  However, this model still requires the growth of $2-3 \mearth$ cores by a different mechanism.  Our model builds on this idea by allowing embryos to grow into giant planet cores as they migrate inward at different speeds.  

\section{Methods}\label{sec:methods}

Our simulations included the effect of a 1-D gaseous protoplanetary disk onto an N-body integrator.  The surface density of the disk was assumed to follow a power law (\reffig{fig:density-profile}).  The temperature structure and other disk parameters were calculated by balancing viscous heating, irradiation, and radiative cooling. No radial diffusive process is taken into account for the equilibrium. However, the scale height and opacity are computed consistently.

Within the context of this disk model we include three effects: eccentricity damping, inclination damping, and Type I migration. 

\subsection{Disk model}

The disk has a power-law surface density profile: 
\begin{align}
\Sigma(R) &= \Sigma(R_0) * \left(\frac{R}{R_0}\right)^{-1/2},
\end{align}
where $\Sigma(R_0=1\unit{AU}) = 300\unit{g/cm^2}$. \modif{The mass of the central star is $M_\star=1\unit{M_\odot}$.}

The disk extends from its inner edge at $0.1\unit{AU}$ out to $100\unit{AU}$. Outside this range, planets feel only gravity.   At its inner edge the disk is smoothed so that the surface density at the inner edge is always $5\unit{g/cm^2}$. The smoothing function is an hyperbolic tangent, the characteristic length is $0.005\unit{AU}$ (scale height at $0.1\unit{AU}$).

Our disk corresponds to a standard alpha disk~\citep{shakura73,pringle1981accretion} with $\alpha=5 \times 10^{-3}$, typical for evolved protoplanetary disk \citep[Fig. 16]{guilloteau2011dual}. 

\begin{figure*}[htb]
\centering
\subfloat[Surface density profile, initial assumption of our disk model.]{\label{fig:density-profile}\includegraphics[width=0.45\textwidth]{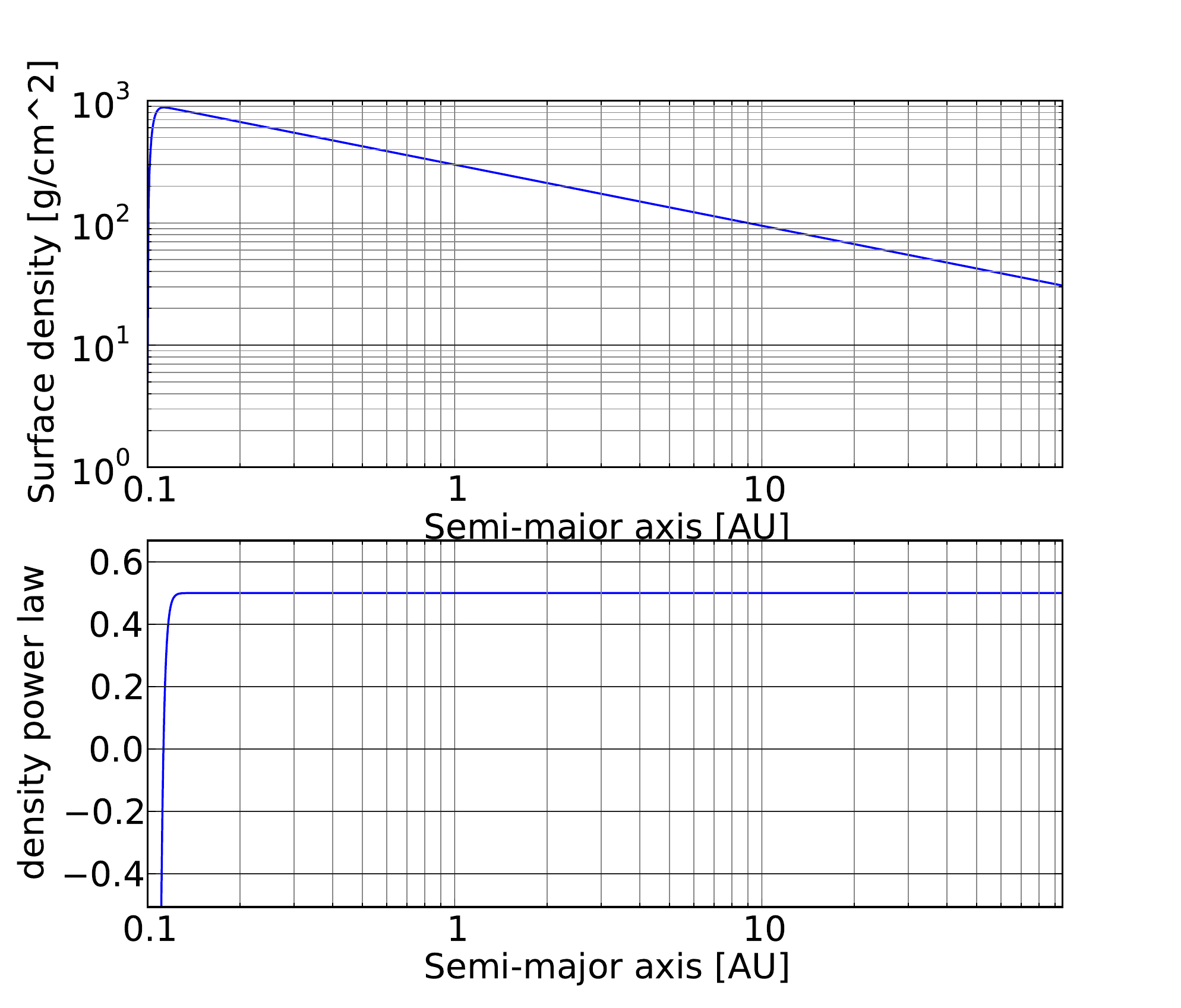}}\hfill
\subfloat[Temperature profile calculated from equilibrium between irradiation, viscous heating and radiative cooling.  The snow line is located at T$\sim$150 K.]{\label{fig:temperature-profile}\includegraphics[width=0.45\textwidth]{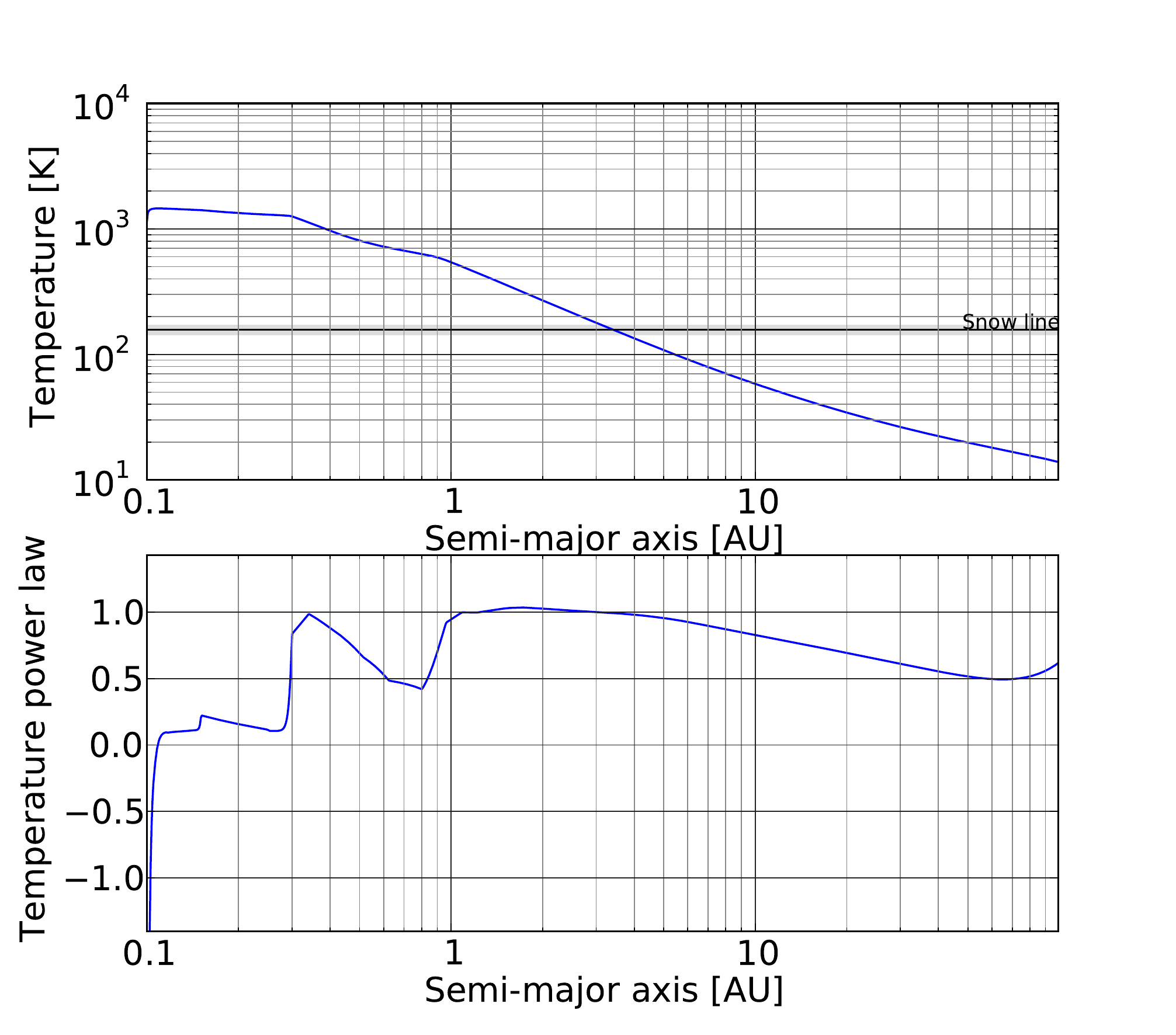}}
\caption{Surface density and temperature profiles from our fiducial disk. The radial extent of our disk is $[0.1;100]\unit{AU}$.}\label{fig:temp_dens_profiles}
\end{figure*}

\begin{figure}[htb]
\centering
\includegraphics[width=\linewidth]{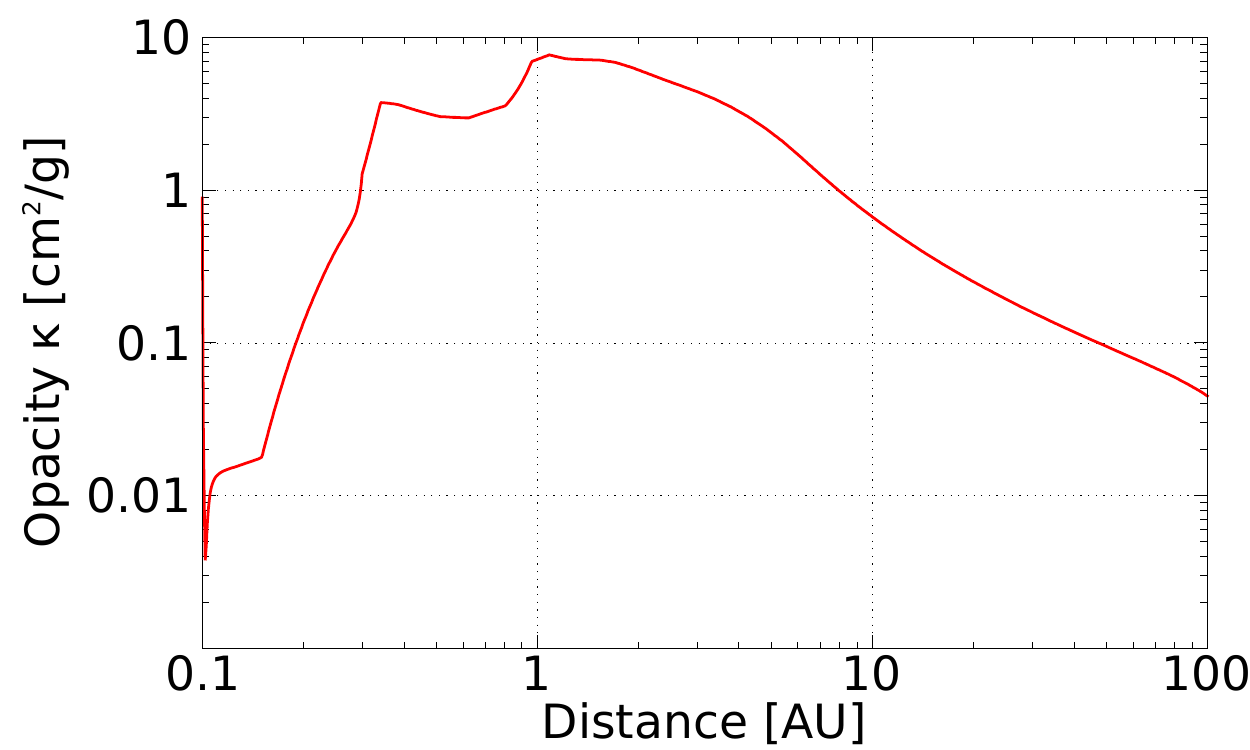}
\caption{Evolution of the disk opacity depending on the temperature and density at each position in the disk.}\label{fig:opacity}
\end{figure}

In the calculation of the disk temperature profile we used the opacity table from \cite{hure2000transition}.  The disk is assumed to be irradiated by a typical TTauri star, with $T_\star = 4000\unit{K}$ and $R_\star = 2.5\unit{R_\odot}$ \citep[Table 2]{hartigan1995disk}.  The disk albedo is assumed to be $\varepsilon=0.5$ (as in \cite{menou2004low}).

To get the temperature \reffig{fig:temperature-profile}, we solve the following equation: 
\begin{align}
0 &= C_\text{bb} + H_\text{en} + H_\text{irr} + H_\text{vis}\nonumber\\
\begin{split}
0 &= - 2\sigma \frac{T^4}{\frac{3}{8}\tau + \frac{\sqrt{3}}{4} + \frac{1}{4\tau}} + 2 \sigma {T_\text{en}}^4 \\
&+ 2 \sigma {T_\star}^4 \frac{{R_\star}^2}{r^2} (1-\varepsilon)
* \left[0.4 \frac{R_\star}{r} + r \od{}{r}\left(\frac{H}{r}\right)\right] + \frac{9}{4} \nu\Sigma\Omega^2\label{eq:equation_energie}
\end{split}
\end{align}
where $C_\text{bb}$ is the radiative cooling through black body radiation of the two surfaces of the disk. $H_\text{en}$ is the heating by the envelope of the disk, assumed to be at $T_\text{en}=10\unit{K}$. $H_\text{irr}$ is the irradiation of the central star (passive disk). $H_\text{vis}$ is the viscous heating of the disk (active disk). \modif{$\nu$ is the kinematic viscosity of the disk. $\Omega$ is the angular velocity of a rotating particle.} $H$ is the scaleheight of the disk and $\tau$ the optical depth. \modif{Regions of the disk were the aspect ratio is decreasing are shadowed, and irradiation is removed in these regions. }

\modif{The temperature profile \reffig{fig:temperature-profile} shows two steep transitions due to opacity transitions \reffig{fig:opacity}.}

\subsection{Eccentricity and inclination damping}\label{sec:ecc_inc_damping}

Planet-disk tidal (type I) interactions tend to rapidly damp a planet's eccentricity and inclination~\citep{papaloizou00,tanaka04,cresswell2006evolution,cresswell07,cresswell2008three}.  We modeled this damping following the prescription of \cite{cresswell2008three}. The characteristic timescale $t_\text{wave}$, defined as follows \cite[eq. (9)]{cresswell2008three}: 
\begin{align}
t_\text{wave} &= \frac{M_\star}{m_p}\frac{M_\star}{\Sigma_p {a_p}^2}\left(\frac{H}{r}\right)^4{\Omega_p}^{-1},
\end{align}
where $\Omega$ is the planet's Keplerian frequency and $\Sigma_p$ is the {\em local} disk surface density.  Based on this, the timescales for eccentricity and inclination damping, $t_e$ and $t_i$, respectively, are defined as \cite[eq. (11) and (12)]{cresswell2008three}:
\begin{subequations}
\begin{align}
\begin{split}
t_e &= \frac{t_\text{wave}}{0.780}\\
&*\left[1-0.14\left(\frac{e}{H/r}\right)^2 + 0.06 \left(\frac{e}{H/r}\right)^3 + 0.18\left(\frac{e}{H/r}\right)\left(\frac{I}{H/r}\right)^2\right]
\end{split}\\
\begin{split}
t_I &= \frac{t_\text{wave}}{0.544}\\
&*\left[1-0.30\left(\frac{I}{H/r}\right)^2 + 0.24 \left(\frac{I}{H/r}\right)^3 + 0.14\left(\frac{e}{H/r}\right)^2\left(\frac{I}{H/r}\right)\right]
\end{split}
\end{align}
\end{subequations}

In practice, we implement the eccentricity and inclination damping by applying artificial accelerations~\citep[eq. (15) and (16)]{cresswell2008three} : 
\begin{subequations}
\begin{align}
\vect{a_e} &= -2 \frac{(\vect{v}.\vect{r})}{r^2 t_e}\vect{r}\\
\vect{a_i} &= - \frac{\vect{v_z}}{t_i}\hat{e}_z
\end{align}
\end{subequations}

To avoid settling into a 2-dimensional setting, inclination damping was dis-activated when $I<0.03^\circ$. \modif{This value was chosen to correspond to the angular size of a roughly Earth-sized planet close to the inner edge of the disk at 0.1 AU, to guarantee that close encounters were not restricted to 2 dimensions.  We tested a few different values of this threshold for inclination damping and saw no effect on the outcome as long as the value was not exceedingly small.  This threshold is nonetheless important to avoid the artificial 2-dimensional regime in which the collision rate is artificially enhanced\citep{chambers01_2d}.  }

\subsection{Type I migration}\label{sec:type_I_migration}
Tidal interactions with the protoplanetary disk torque a planet's orbit and cause the orbit to shrink or grow.  The torques can be divided into the differential Lindblad torque, the unsaturated and saturated corotation torque. The net effect of the sum of these torques is what is called type I migration.  We modeled type I migration using the prescription of \cite{paardekooper2010torque,paardekooper2011torque}. 

The following torque will be named as follow: 
\begin{itemize}
\item \textbf{c} correspond to Corotation torque, while \textbf{L} is for Lindblad ;
\item \textbf{hs} is for horseshoe drag, when the torque is fully unsaturated ; 
\item \textbf{lin} is for the linear corotation torque ;
\item \textbf{baro} is for the barotropic part of the corotation torque ;
\item \textbf{ent} is for the entropy related part of the corotation torque.
\end{itemize}
\modif{The positive part of the corotation torque has to compensate for the negative Lindblad torque}. The corotation torque has two contribution, a linear and a horseshoe related part. Each one of them has two parts, an entropy related and a barotropic one.

The differential Lindblad torque is given by \citep[eq. (14)]{paardekooper2010torque} : 
\begin{align}
\gamma \Gamma_L/\Gamma_0 &= - \left(2.5 +1.7\beta -0.1d\right) \left(\frac{0.4}{b/h}\right)^{0.71}\label{eq:lindblad-torque}
\end{align}
where $\gamma$ is the adiabatic index, $b/h$ the smoothing parameter \modif{of the planet's gravitational potential in units of its Hill radius}, and $\beta$ and $d$ are the negative indexes of temperature ($T \propto R^{-\beta}$) and surface density power laws ($\Sigma \propto R^{-d}$). 

The non-saturated part of the corotation torque is given by \citep[eq. (45)]{paardekooper2010torque} :
\begin{subequations}
\begin{align}
\gamma \Gamma_\text{c,hs,baro}/\Gamma_0 &= 1.1\left( \frac{3}{2} - d\right)\left(\frac{0.4}{b/h}\right)\\
\gamma \Gamma_\text{c,hs,ent}/\Gamma_0 &= \frac{\xi}{\gamma}\left(10.1\sqrt{\frac{0.4}{b/h}} - 2.2\right)\left(\frac{0.4}{b/h}\right)
\end{align}\label{eq:saturated-corotation-torque}
\end{subequations}
and the linear ones \citep[eq. (17)]{paardekooper2010torque} :
\begin{subequations}
\begin{align}
\gamma \Gamma_\text{c,lin,baro}/\Gamma_0 &= 0.7\left( \frac{3}{2} - d\right)\left(\frac{0.4}{b/h}\right)^{1.26}\\
\gamma \Gamma_\text{c,lin,ent}/\Gamma_0 &= \xi\left[2.2\left(\frac{0.4}{b/h}\right)^{0.71} - \frac{1.4}{\gamma}\left(\frac{0.4}{b/h}\right)^{1.26}\right]
\end{align}\label{eq:linear-corotation-torque}
\end{subequations}
$\xi=\beta - (\gamma-1)d$ is the negative index of the entropy power law ($S \propto R^{-\xi}$).

Theses torques are normalized by a reference torque $\Gamma_0$, defined as: 
\begin{align}
\Gamma_0 &= \left(\frac{q}{h}\right)^2\Sigma_p {r_p}^4 {\Omega_p}^2
\end{align}

We also include the damping of the corotation torque for planets on eccentric orbits. The formula used is given in \citep{cossou2013convergence} : 
\begin{equation}
D = \frac{\Gamma_C(e)}{\Gamma_C (e=0)} = 1 + a \cdot \left[\tanh(c) - \tanh\left(\frac{b * e}{x_s}+c\right)\right],\label{eq:eccentricity-influence}
\end{equation}
where $x_s$ represents the half-width of the horseshoe region divided by the semimajor axis, $e$ is the planet's eccentricity, and our fit produced $a = 0.45$, $b = 3.46$, and $c = -2.34$.

\begin{figure}[htb]
\centering
\includegraphics[width=\linewidth]{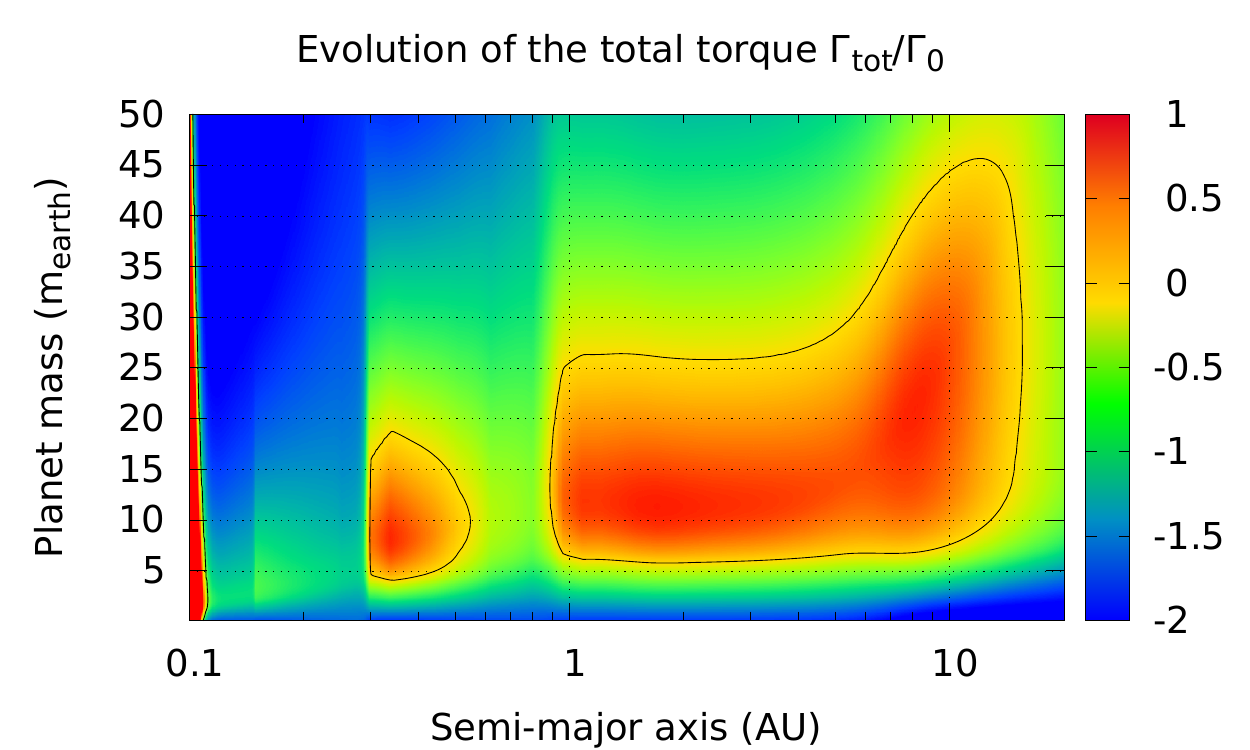}
\caption{Type I migration map for our disk model.  The color indicates the direction and strength of migration. Zero-migration contours enclose the region of outward migration.  This map assumes circular planetary orbits. }\label{fig:migration_map}
\end{figure}

Figure~\ref{fig:migration_map} shows the type I migration map for our disk model.  Despite the relatively simple assumptions made in our disk model, the map shows complex structure.  The main region of outward migration extends from roughly 1 to 15 AU, and is confined to planets with masses of 5 to $25 \mearth$.  There is an additional smaller zone of outward migration from 0.3-0.6 AU.  The inner edge of the disk represents a very strong barrier to inward migration, where a very strong positive corotation torque is generated by the strong density and temperature gradients~\citep{masset06}.  This map is similar to those calculated by other authors~\citep{kretke12,bitsch13,bitsch2014stellar}.  Our calculation is robust: we verified that, for the same assumptions, our procedure reproduces the migration maps of \cite{bitsch13}.

\subsection{Simulations}
Our simulations started with a population of planetary embryos that are assumed to have formed in a prior phase of growth, presumably by accretion of planetesimals~\citep{thommes03,levison10} or pebbles~\citep{lambrechts12,morby12}.  Although this is an area of active research, there are indeed models that predict that planetesimals should form first at larger orbital radii~\citep{chambers10}.

For a given disk mass, \modif{embryos were assigned individual masses uniformly distributed between 0.1 and $2\mearth$.  The total mass in embryos remained constant for all simulations in a given set}.  Embryos were placed starting from $1\unit{AU}$. The separation of two consecutive embryos was calculated so that each embryo contained the appropriate dust mass, following our disk surface density profile and assuming a 100:1 gas-to-dust ratio. The last embryo is around 20 AU. Embryos started with orbital eccentricities of $10^{-5}<e<10^{-3}$ and inclinations of $-1^\circ<I<1^\circ$.

We performed multiple sets of simulations. \modif{The initial properties of the embryos (masses and orbits) were drawn from the same underlying distribution for each of the 100 simulations in a given set, but with a different random number seed.}  First, we performed simulations in which the disk did not evolve.  These systems were integrated for 3 Myr.  We ran sets of simulations with total mass in embryos of $21$, $42$ and $84\unit{M_\oplus}$.  This essentially corresponds to a factor of 2 increase/decrease in the dust-to-gas ratio, i.e., the metallicity.

Second, we performed simulations that included dispersal of the gaseous disk. The typical disk lifetime is thought to be a few Myr~\citep{haisch01,briceno2001CIDA,lada2006spitzer,hillenbrand08}.  We therefore modeled disk dissipation as an exponential decay with two different timescales.  For the first 3 Myr of the disk lifetime the disk surface density decreased uniformly at all radii as $\Sigma = \Sigma_0 \exp\left(-t/t_1\right)$, where $t_1$ = 2 Myr.  After that point the disk was rapidly dissipated with a second exponential with a timescale of $t_2$ = 50,000 years. This second, fast decay was designed to capture the approximate effect of the fast final dissipation of the disk by photo-evaporation~\citep{wolk1996search,simon1995disk,alexander06,currie09,owen2010radiation}.  The total embryo mass in these simulations was $42\unit{M_\oplus}$.  

\subsection{Integrator}
Our code is based on the hybrid {\tt Mercury} \citep{chambers1999hybrid} integrator.  Particles were assumed to hit the central star inside a radius of 0.005 AU and were assumed to be ejected from the system outside a radius of 100 AU.  Each simulation was integrated for 3 to 100 Myr.  We used a timestep of 0.4 days.  We carefully checked that integration errors were negligible ($dE/E < 10^{-6}$) in to the assumed inner edge ($0.1\unit{AU}$)~\citep[numerical tests based on Appendix A of][]{raymond11}.  Of course, given the damping from the disk, the total energy of the system was not conserved.  Collisions between bodies were assumed to be inelastic and to conserve linear momentum.  

\section{Static disk simulations}\label{sec:static}

We now discuss the results of our simulations in non-evolving, static disks, for 3 Myr.  We first focus on planets that migrated to the inner edge of the disk (Section~\ref{sec:static_HSE}).  We call these hot super-Earths.  Even though many may correspond more closely with mini-Neptunes, we group those two categories for simplicity.  In the following section we discuss planets that may represent giant planet cores (Section~\ref{sec:static_GPC}).  

\subsection{Hot super-Earths}\label{sec:static_HSE}
In our model hot super-Earths form when embryos migrate inward but do not reach the critical mass to reverse their migration. This mechanism, proposed by \cite{terquem2007migration}, consists of the migration of embryos across large orbital distances to the inner edge of the disk.  Close to the star, collisions between embryos grow compact systems of planets. The main difference here, compared to \cite{terquem2007migration}, is that systematic inward migration is not assumed but rather computed consistently. 

\begin{figure}[htb]
 \centering
 \includegraphics[width=0.85\linewidth]{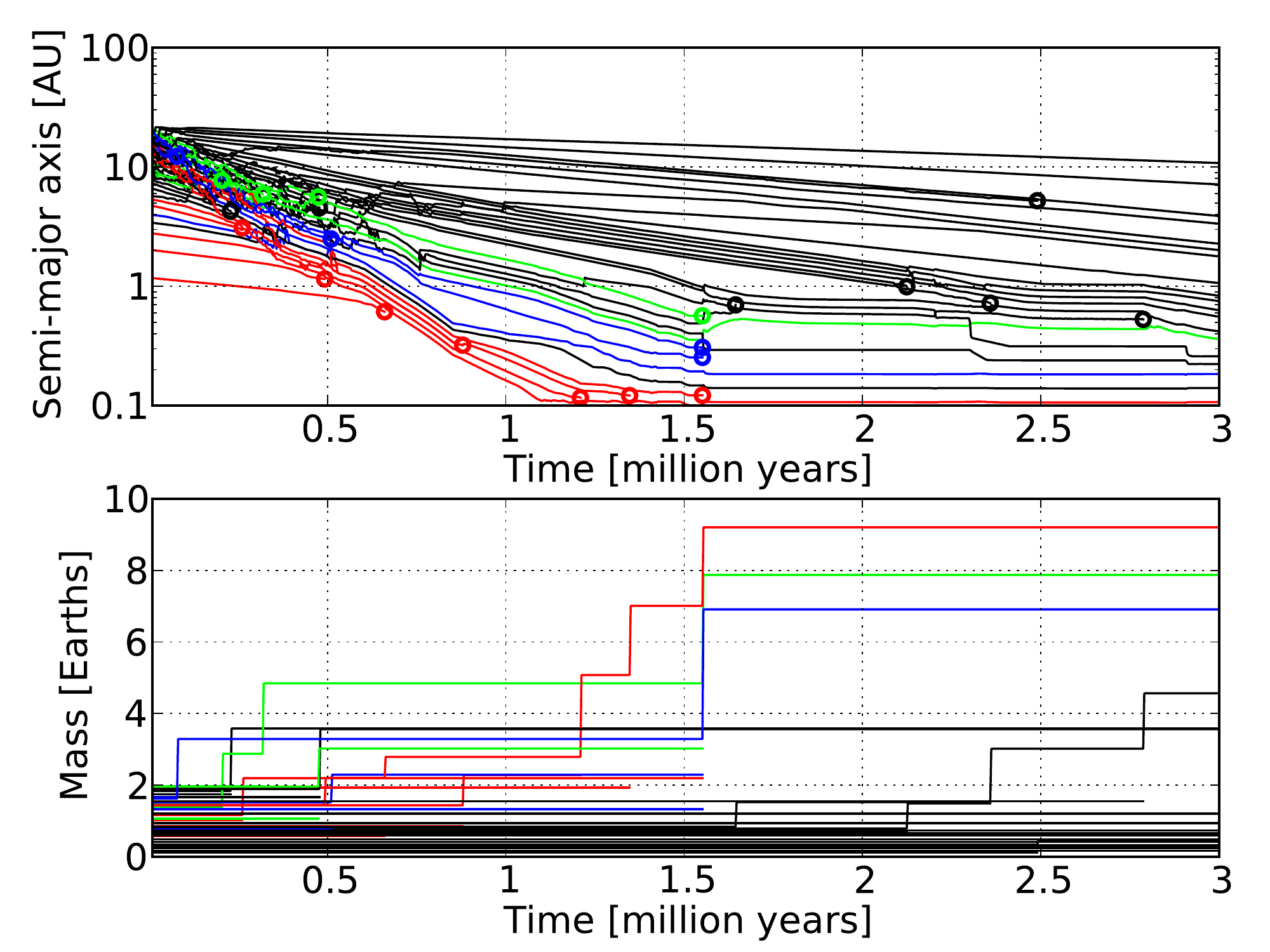}
 \caption{Orbital evolution (top panel) and growth (bottom panel) of planetary embryos in a simulation that produced a system of hot super-Earths. Each circle represents a collision. Red, green, and blue lines trace the evolution of the embryos that collide to form the three more massive final planets.  All other embryos are shown in black.  Embryos migrate inward toward the inner edge of the disk at 0.1 AU, where collisions are enhanced due to strong convergence.  The inner edge stops embryos from migrating further in (although this is not always the case; see below).}\label{fig:HSE_case}
\end{figure}

Figure~\ref{fig:HSE_case} shows the evolution of a simulation that formed a system of hot super-Earths.  Embryos starting between 1 and 20 AU migrated inward together but at different speeds given the strong mass dependence of the type I migration rate.  Multiple accretion events occurred during migration at orbital radii of several AU.  More collisions between embryos occurred in the inner parts of the disk, as migrating embryos piled into a resonant chain.  This chain balanced the cumulative negative torques felt by many embryos with the very strong positive torque felt by the innermost embryo.  When the resonant chain included too many embryos, instabilities between the orbits of adjacent embryos caused collisions.  Further migration produced a new resonant chain and the process continued.  This same mechanism has been identified in previous work~\citep[see, for example, ][]{morby08,sandor11}.  

During the simulation from \reffig{fig:HSE_case}, two planets formed with masses above the critical mass for outward migration (see migration map in \reffig{fig:migration_map}).  These planets only reached such large masses once they were too close-in for long-range outward migration.  An outward-migrating planet can be pushed back inward by inward-migrating ones by either overwhelming the torque balance or by exciting the eccentricity of the large planet and thereby weakening its positive corotation torque~\citep[][; see also Section \ref{sec:dissip_GPC}]{cossou2013convergence}.  

At the end of simulation from \reffig{fig:HSE_case} a system of 7 planets more massive than $1 \mearth$ survived inside 0.5 AU.  An extended population of embryos with masses of a few tenths of an Earth mass survived out to 10 AU and slowly migrated inward in the last millions years, when the dissipation of the disk is expected to slow then stop migration.  This extended population comprised a total of $6.2 \mearth$, just 15\% of the initial mass budget in the outer disk.  

In the static disk simulations compact systems of hot super-Earths formed systematically.  They grew from planetary embryos that migrated inward and collided, and piled up at the inner edge of the disk in compact orbital configurations.  The planets' orbits were locked in chains of mean motion resonances.  This is a natural outcome of convergent migration in the dissipative environments characteristic of gaseous protoplanetary disks~\citep{snellgrove01,lee02,pierens13}.  

\begin{figure}[htb]
 \centering
 \includegraphics[width=0.85\linewidth]{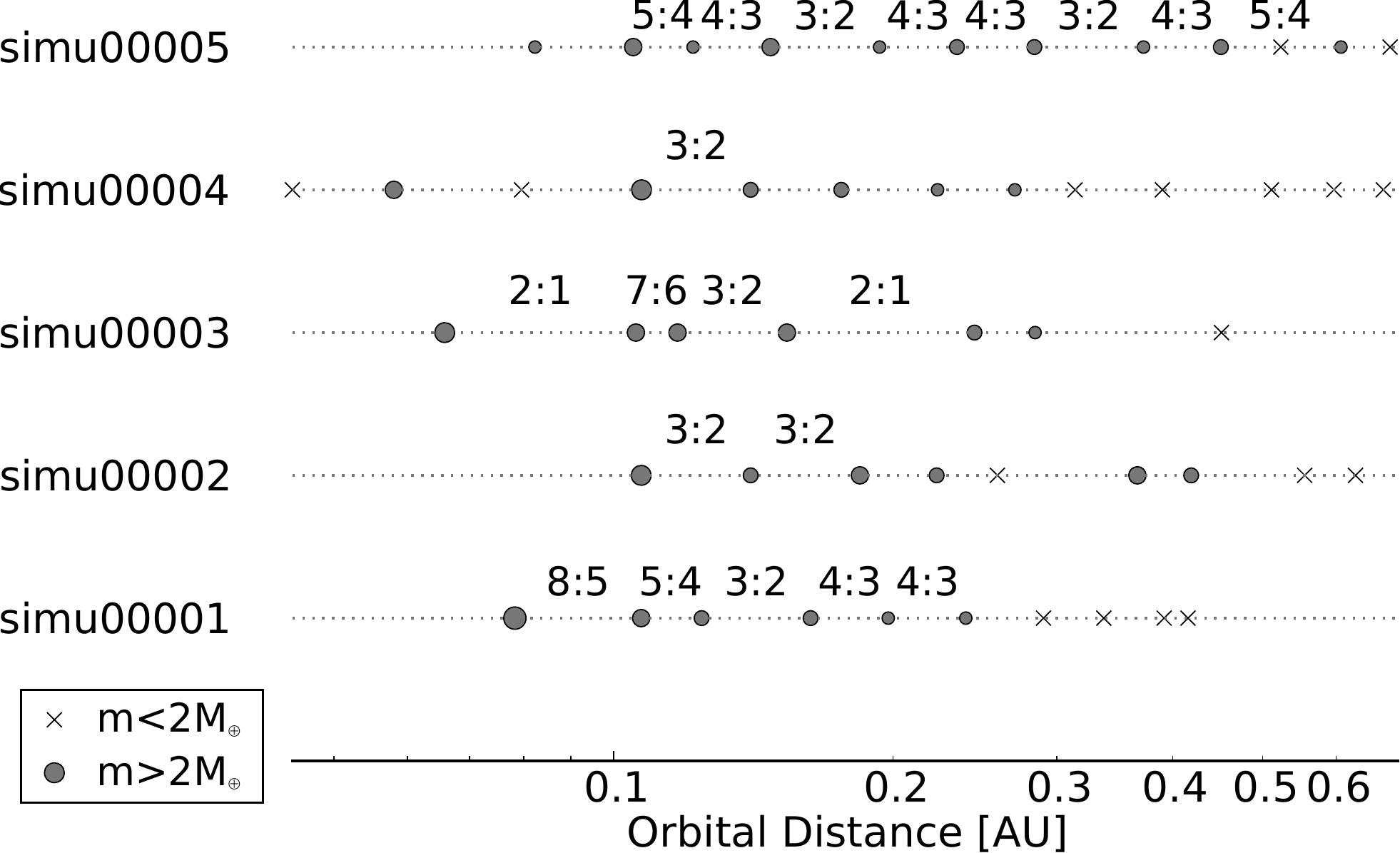}
 \caption{Final configuration of five systems after 3 million years of evolution in a static disk. The inner edge of the disk is located at 0.1 AU.  The system in \reffig{fig:HSE_case} is {\em simu00002}.}\label{fig:5_inner_systems}
\end{figure}

\reffig{fig:5_inner_systems} shows the final orbital configuration of the inner parts of five systems formed in the static disk simulations (including the one from \reffig{fig:HSE_case}, which is "simu00002").  The resonances between adjacent pairs of planets that were found by an automated search are labeled.  As expected, most adjacent planets are indeed in resonance.  However, in some cases our script did not find resonances because the libration amplitude of the associated resonant angles was larger than our arbitrarily-chosen critical value of $50^\circ$ (standard deviation less than $50^\circ$ for the last 150,000 yr).  Systems with larger libration amplitudes are still resonant -- basically all of the planets in these simulations are in resonance -- but are less deep in the resonance.  Pairs of planets in ``shallower'' resonances are interesting because they are the least stable; without gas many such resonances should quickly become unstable and likely trigger a later phase of scattering and accretion~\citep[see, e.g.,][]{iwasaki01,iwasaki06,matsumoto12}.  

The innermost planet is usually interior to 0.1 AU, the inner edge of our model disk (\reffig{fig:5_inner_systems}).  These inner planets were pushed over the edge by the swarm of migrating embryos and stranded in the gas-free inner cavity of the disk.  The innermost planets are therefore often not in resonance with the planets immediately exterior to their orbits.  However, more distant planets that formed within the gas disk are almost always in long resonant chains.  

The configuration of the resonant chain depends on the migration speed.  This is primarily a function of the disk and planet masses (see Section 2.3).  The faster the migration speed, the stronger the convergent migration and the more compact the resonant chain.  At the same time, the faster the migration, the more accretion during this process.  For these reasons it is easier to generate a very wide resonant chain in a lower-mass disk and with lower-mass planets.  This issue is important for matching observations and will return in the discussion (see Section 6.3).  

\begin{figure}[htb]
 \centering
 \includegraphics[width=0.85\linewidth]{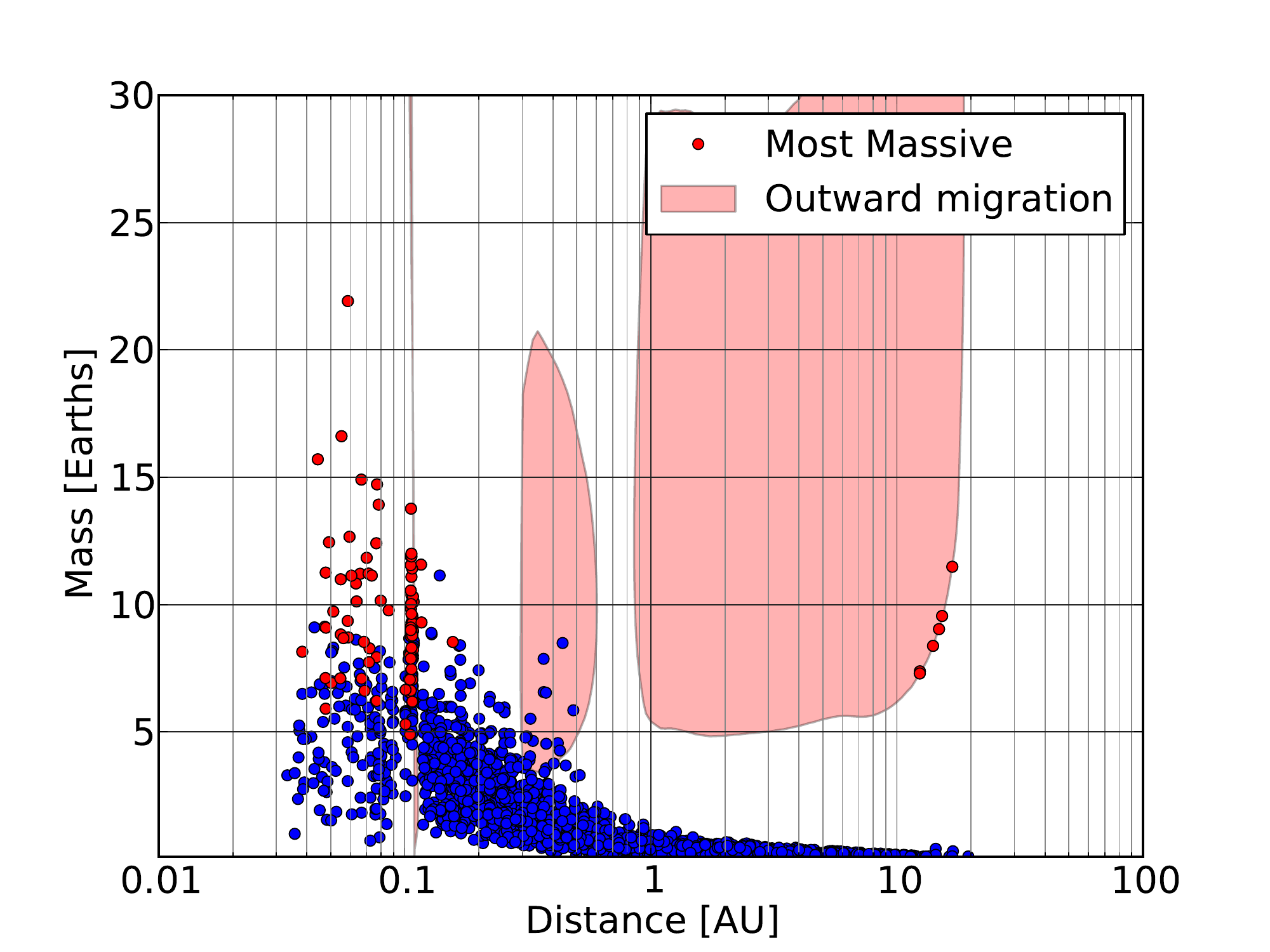}
 \caption{Mass vs. semimajor axis for the planets that formed in the 100 static disk simulations after 3 Myr of evolution. Each red dot represents the most massive planet in a given system.}\label{fig:fiducial_static_stat}
\end{figure}

Figure~\ref{fig:fiducial_static_stat} shows the final orbital separation and mass of the planets that formed in the simulations with static disks.  A population of planets -- often including the most massive planet in a given system -- exists as close in as 0.034 AU, far interior to the inner edge of the disk.  There is a pileup of planets at 0.1 AU, a signature of the migration trap at the disk inner edge.  Planets with comparable masses of up to $5-10 \mearth$ exist out to roughly 0.5 AU.  Farther out is a population of low-mass planetary embryos that extends to 20 AU.  Accretion remains unfinished among these bodies.  Only giant planet cores have large masses at large orbital radii (see Section \ref{sec:static_GPC} below).

\subsubsection{Comparison with observations}\label{sec:static_comp_obs}

We now compare our simulated hot super-Earths with observations.  Our comparison sample is based on recent statistical analyses of the Kepler planet candidates~\citep[e.g.][]{youdin11,tremaine12}.  We adopt the radius and period constraints from \citep{fang2012architecture} to define our subset of comparison: $m>3.3\mearth$\footnote{Transforming the $1.5R_\oplus$ threshold assuming Earth's bulk density for all simulated planets.} and $a<0.66\unit{AU}$\footnote{Transforming the $200\unit{days}$ threshold assuming the mass of the sun for all the simulated stars.}. The assumed eccentricity distribution is taken as a Raleigh distribution with $\overline{e}\sim 0.1$\citep{moorhead2011distribution}.  The assumed underlying distribution of mutual inclinations is a Gaussian with zero mean and a standard deviation of $\sigma_I\sim 1.5^\circ$.  This puts 85\% of the planets below $3^\circ$ (as noted in \cite{fang2012architecture}).  The distribution of period ratios between adjacent planets are taken from known exoplanet systems.

\begin{figure*}[htb]
\centering
\subfloat[Eccentricities $e$]{\includegraphics[width=0.45\textwidth]{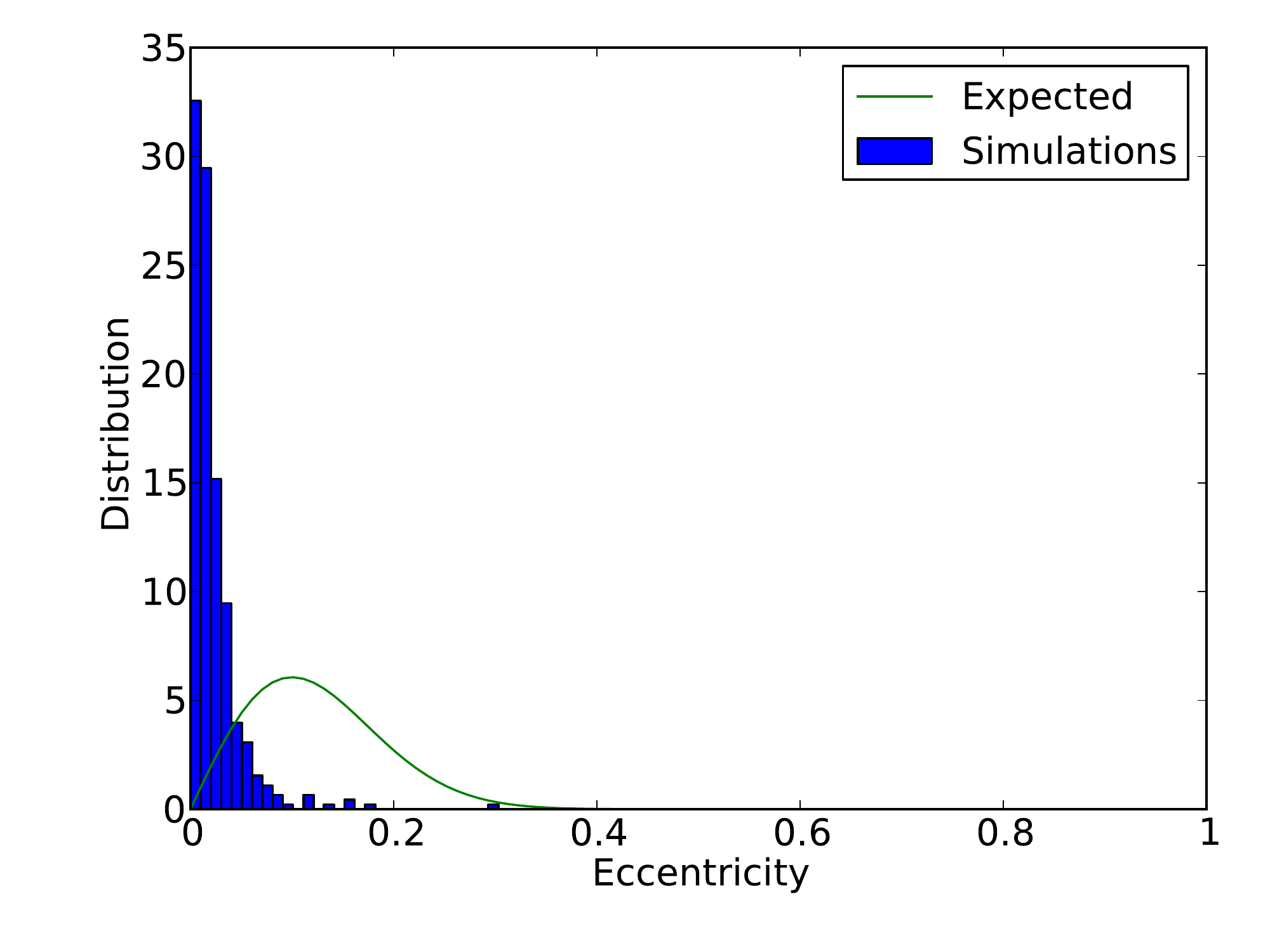}}\hfill
\subfloat[Mutual inclinations $\Delta I$]{\includegraphics[width=0.45\textwidth]{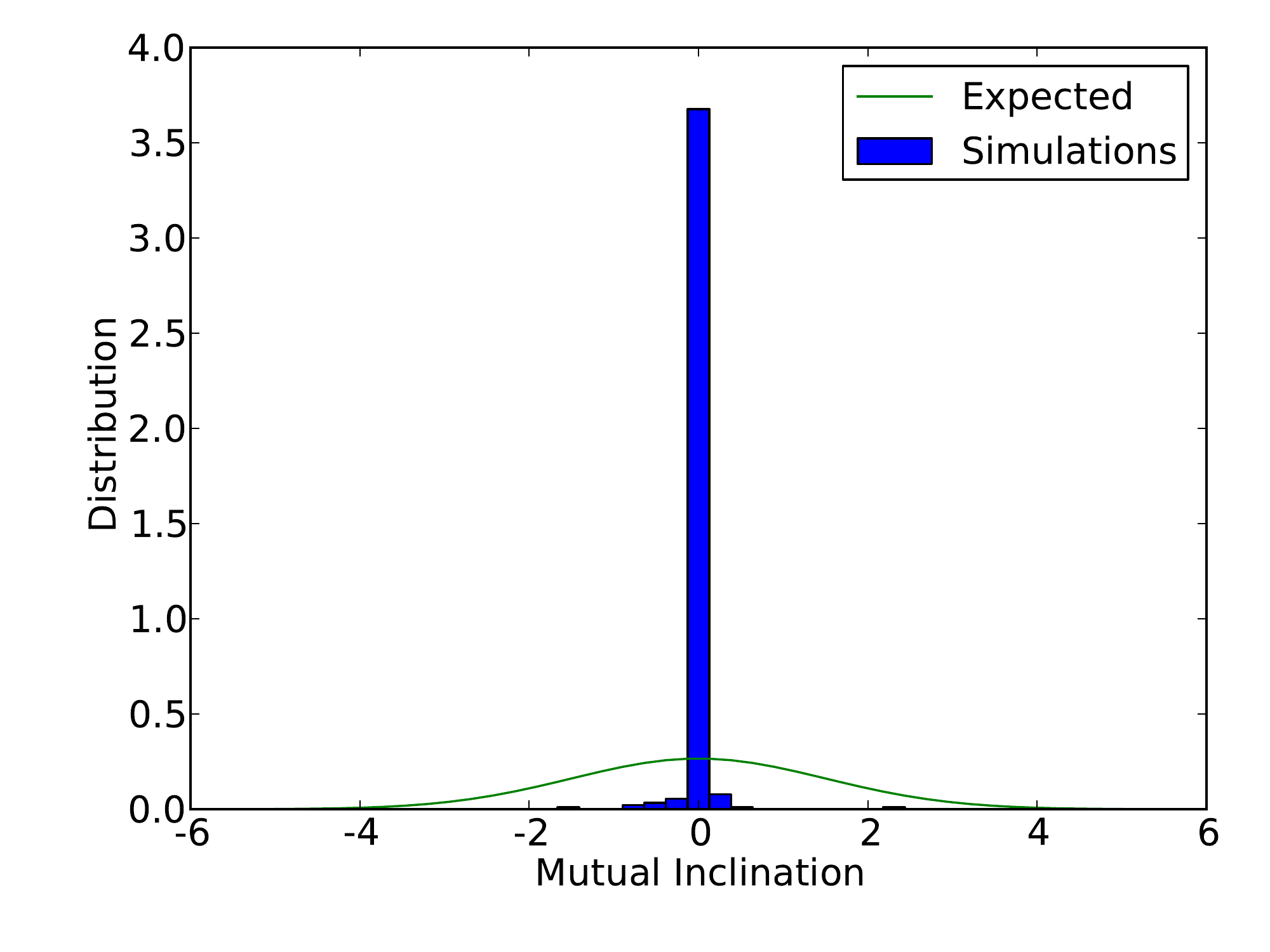}}

\subfloat[Period ratios between adjacent planets, no selection effects (except $a<1\unit{AU}$)]{\includegraphics[width=0.45\textwidth]{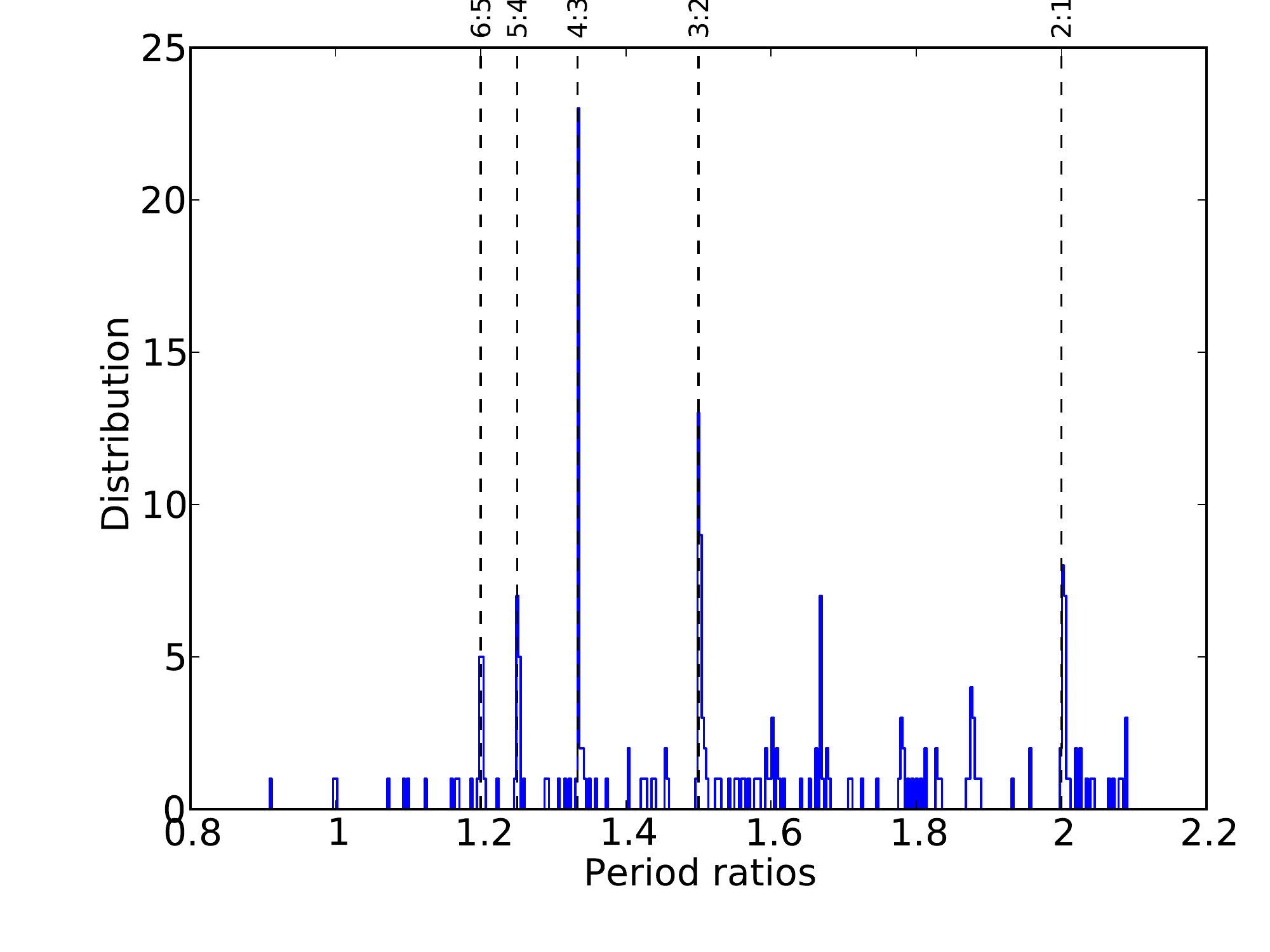}}\hfill
\subfloat[Cumulative distribution of the period ratios between adjacent planets, including selection effects]{\includegraphics[width=0.45\textwidth]{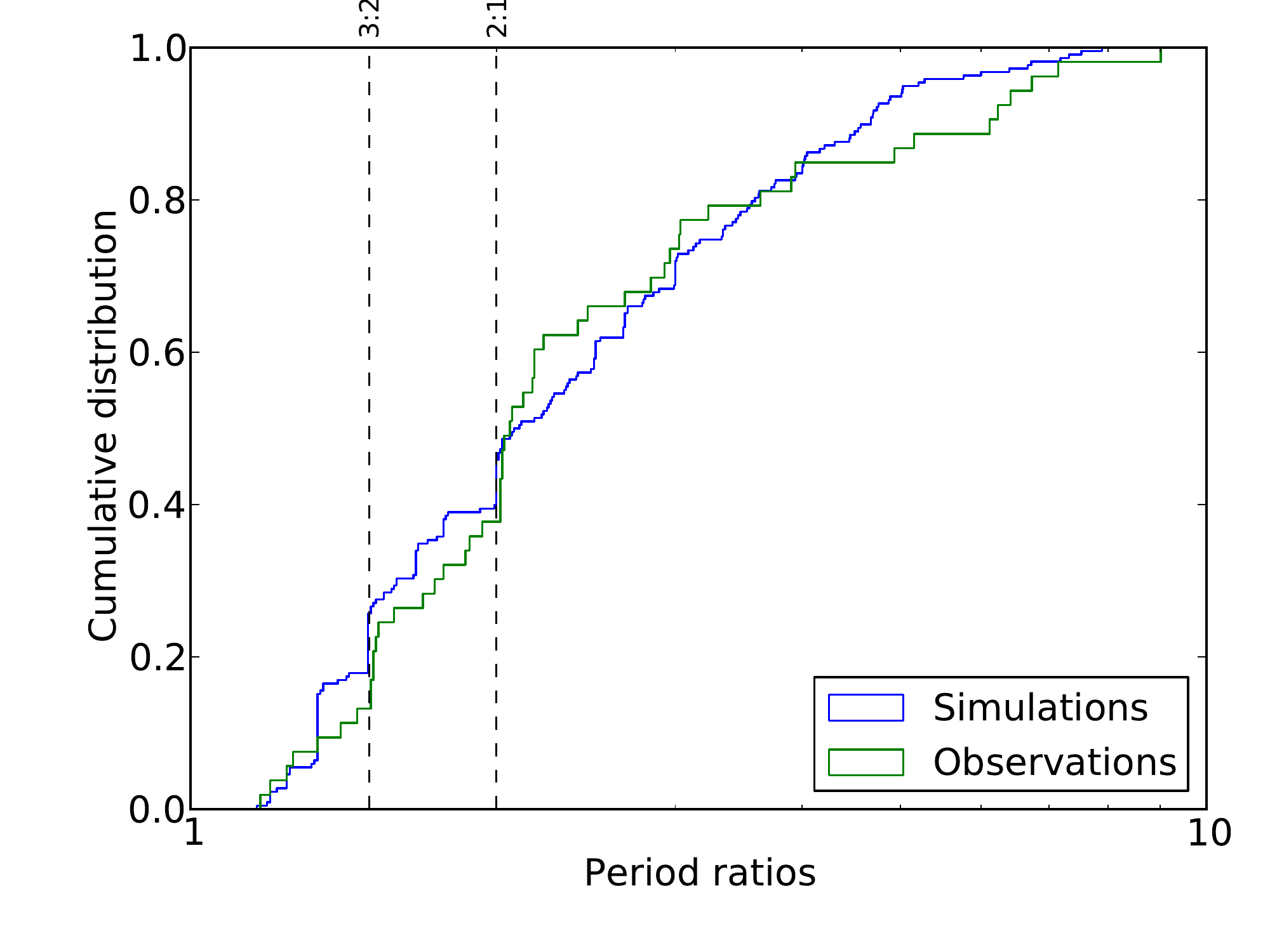}}
\caption{Comparison between the planets that formed in the simulations with static disks and observed extra-solar planets.  Both samples are limited to the range $R>1.5R_\oplus$ and $p<200\unit{days}$, except for the non cumulative histogram of simulation only (without observations) that shows that the vast majority of hot super Earths are in resonance.}\label{fig:comp_no_dissip}
\end{figure*}

Figure~\ref{fig:comp_no_dissip} compares the simulated {\em detectable} hot super-Earths with the comparison sample.  The figure compares the planets' orbital eccentricities and inclinations and the period ratios of adjacent planets.  The simulated systems are much dynamically colder than the observed ones.  The simulated planets are clustered at very low eccentricities and inclinations, with only a small fraction of planets extending even to $e \gtrsim 0.1$ or $i \gtrsim 0.5^\circ$.  This is a simple consequence of the very dissipative environment of our modeled disk, which quickly damps eccentricities and inclinations (see Section \ref{sec:ecc_inc_damping}).  

The simulated planets are almost entirely locked in mean motion resonances.  This is clear from the shark spikes in the distribution of period ratios in \reffig{fig:comp_no_dissip}. In contrast, when we apply the appropriate selection biases, the period ratios of the simulated planets provide an excellent match to the observed Kepler candidate systems. 


To summarize, simulations in static disks produce planetary systems that are too dynamically cold and too commonly found in resonance.

\subsection{Giant planet cores}\label{sec:static_GPC}

In this model, giant planet cores are represented by embryos that grow massive enough quickly enough that the direction of their migration is reversed.  These cores migrate outward and are generally stabilized at a zero-torque radius represented in Fig. 2 by the right edge of the black contours.  The cores can then remain at large orbital separation for long timescales, with masses that should be conducive to efficient gas accretion~\citep[e.g.,][]{pollack96,hubickyj2005accretion}.

With initial masses between $0.1$ and $2\mearth$, embryos \modif{are not massive enough to maintain unsaturated horseshoe drag}. Instead, their corotation torque saturates because of the U-turn timescale is too short compared with the diffusion timescale~\citep{paardekooper2010torque}.  Within this mass range, no matter their initial position in the disk, embryos migrate inward~\citep{bitsch13,bitsch2014stellar}. However, collisions that occur during migration can in some cases cause a transition into the fully unsaturated horseshoe drag regime that generates outward migration.  

\begin{figure}[htb]
 \centering
 \includegraphics[width=0.85\linewidth]{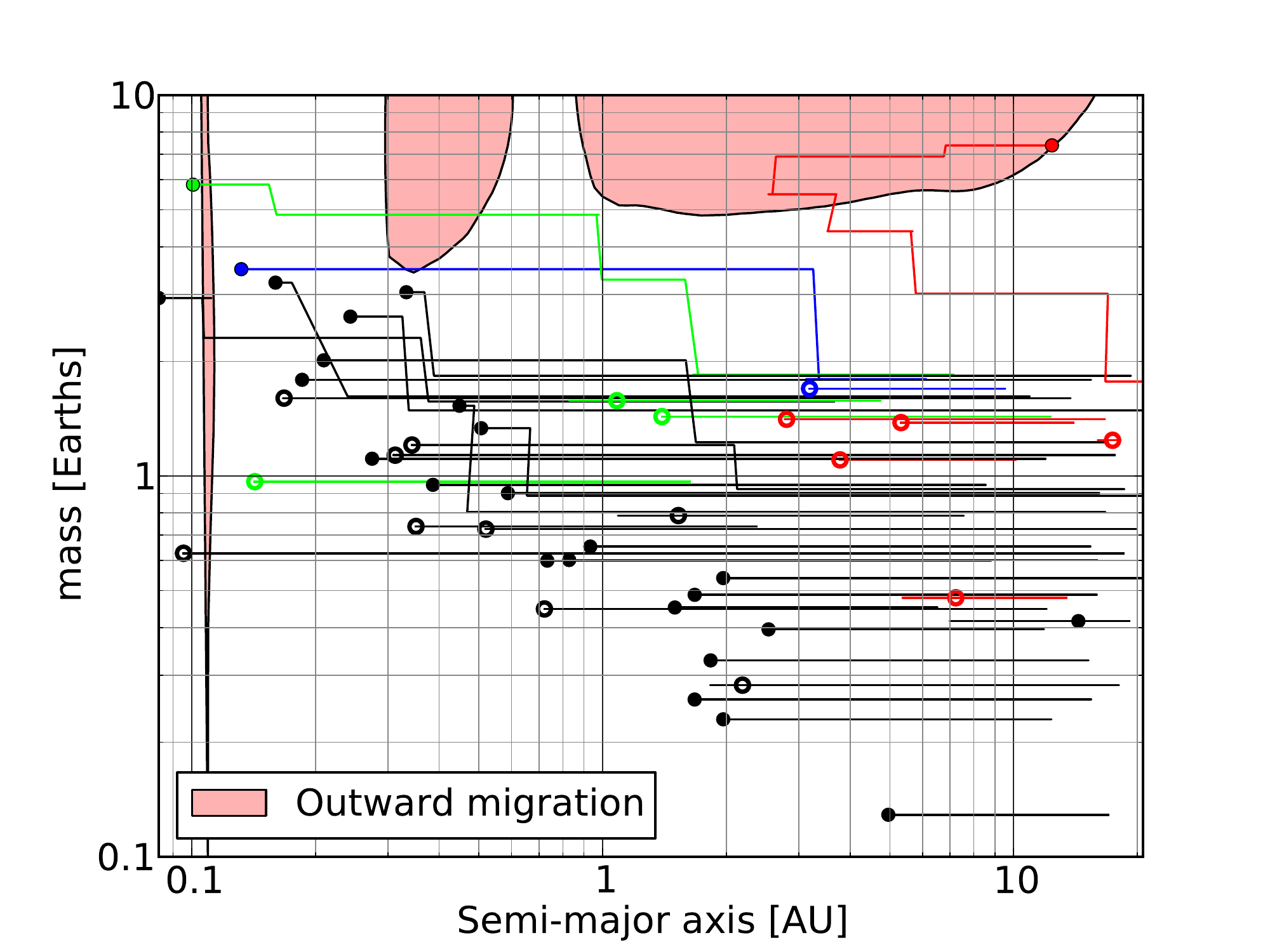}
 \caption{Mass vs. semimajor axis evolution of the embryos in a simulation that produced a giant planet core. Horizontal displacement represents orbital migration, either inward or out- ward. Each vertical shift represents a collision. The three most massive planets are colored in red, green and blue. Each empty circle represents a collision, and each filled circle represents a surviving planet at its final position. Here the most massive embryo (in red) migrated outward to become a giant planet core.}\label{fig:GPC_case}
\end{figure}

\reffig{fig:GPC_case} shows a simulation that produced a giant planet core, a $7.5\mearth$ planet at $12\unit{AU}$.  At $400,000\unit{years}$ a collision between a $4.3\mearth$ and a $1.2\mearth$ embryo created a $5.5\mearth$ body, theoretically massive enough to reverse its migration.  However, the presence of other, nearby embryos prevented it from migrating outward at this point.  These other embryos became locked in exterior mean motion resonances.  This had two consequences.  First, the negative Type I torques acting on those embryos were added to the core's torque budget.  Second, resonant excitation of the core's eccentricity weakened the core's corotation torque~\citep{cossou2013convergence,fendyke2014corotation}.  

The core got a second chance. Another collision at 700,000 years grew the core to $6.9\mearth$.  At this point the core's strong positive corotation torque overwhelmed nearby embryos and the core migrated outward. This demonstrates the important role of core mass: a core must be more massive than its neighboring embryos in order to dominate its vicinity and migrate outward~\citep{cossou2013convergence}.  As the core migrated outward it scattered multiple embryos interior to its orbit.  Two small embryos were trapped in resonance with the migrating core and ended up being pushed outward along with it.

Our fiducial set of simulations in a static disk produced giant planet cores located between 12 and 17 AU with masses of 7 to $12 \mearth$ (see Fig.~\ref{fig:fiducial_static_stat}).  We did not model gas accretion onto growing cores~\citep[see, for example,][]{levison1998modeling,hellary2012global}.  Objects like the one produced in Fig.~\ref{fig:GPC_case} therefore only represent candidate giant planet cores.  Nonetheless, these objects have all of the conditions needed to form full-fledged gas giants.  They are all larger than the estimated lower limit for rapid gas accretion~\citep{ikoma2000formation,hubickyj2005accretion}.  


The cores are also located in a part of the disk that is cold enough that gas accretion should be relatively efficient~\citep{ikoma01}.  In addition, their orbits are relatively stable as the zero-torque radius evolves on a timescale that corresponds to the disk evolution~\citep{lyra10,horn12,bitsch2014stellar}, which is far longer than the standard type I migration timescale.  

Giant planet cores formed in just 5\% (5/100) of the fiducial simulations \modif{\reffig{fig:fiducial_static_stat}}.  This is lower than estimates of the frequency of giant exoplanets~\citep{cumming08,mayor11}.  However, we expect the frequency from our simulations to be an underestimate because additional sources of growth have not been accounted for.  For instance, we did not include pebbles or planetesimals, which can in some cases be efficiently accreted onto growing embryos~\citep{levison10,lambrechts12,morby12}.  We also did not take gas accretion into account.  Even relatively slow gas accretion that increased the masses of the larger embryos by 10-50\% could affect the probability of forming giant planet cores.  In addition, giant impacts between embryos may actually stimulate rapid gas accretion~\citep{broeg2012giant}.  

We performed a simple experiment to test the effect of additional sources of accretion.  We tracked the evolution of each embryo to determine how many crossed the migration-reversal threshold ($\sim6\mearth$ farther than 1.5 AU).  We found that 12\% (12/100) of the simulations have embryos that cross the threshold.  Only 5/12 (42\%) became actual giant planet core candidates since their corotation torques can be attenuated or overwhelmed by outer embryos.  We also tracked embryos that came within 33\% in mass of the migration-reversal threshold ($\sim4\mearth$ at 1.5AU).  \modif{We found that 72\% of simulations had candidates that met this criterion and came close to becoming giant planet cores.  This is clearly a crude tool but it helps gauge the potential of this model.  Of course, the fraction of systems with potential cores represents an upper limit to the efficiency of core formation because if embryos accreted more quickly they would also migrate faster.  In fact, \cite{lega2014migration} found that migration of low-mass planets may be even faster than predicted in our model. } 


\subsubsection{Influence of the solid to gas mass ratio: a crude metallicity tracer}
Higher-metallicity stars have a much higher occurrence rate of giant exoplanets~\citep{gonzalez97,santos2004spectroscopic,fischer05}.  However, there is no observed correlation between the stellar metallicity and the frequency of hot super-Earths~\citep{ghezzi10,buchhave12,mann12}.  

We performed a crude test of the effect of stellar metallicity on our model.  We approximated by assuming that the dust-to-gas ratio in the disk follows the stellar metallicity.  \modif{We neglect the effect of the increased dust mass on the disk opacity and temperature structure.} We ran simulations with the same gas disk model as our fiducial case but with half and twice the mass in embryos.  Given the complex scalings between the type I migration map and the disk parameters, this was a far simpler experiment than changing both the disk and embryo characteristics.  Such experiments were performed during the course of C. Cossou's thesis and are included in his thesis manuscript \citep{cossou2013thesis}, albeit with a slightly older version of the code. 

\begin{figure}[htb]
 \centering
 \includegraphics[width=0.85\linewidth]{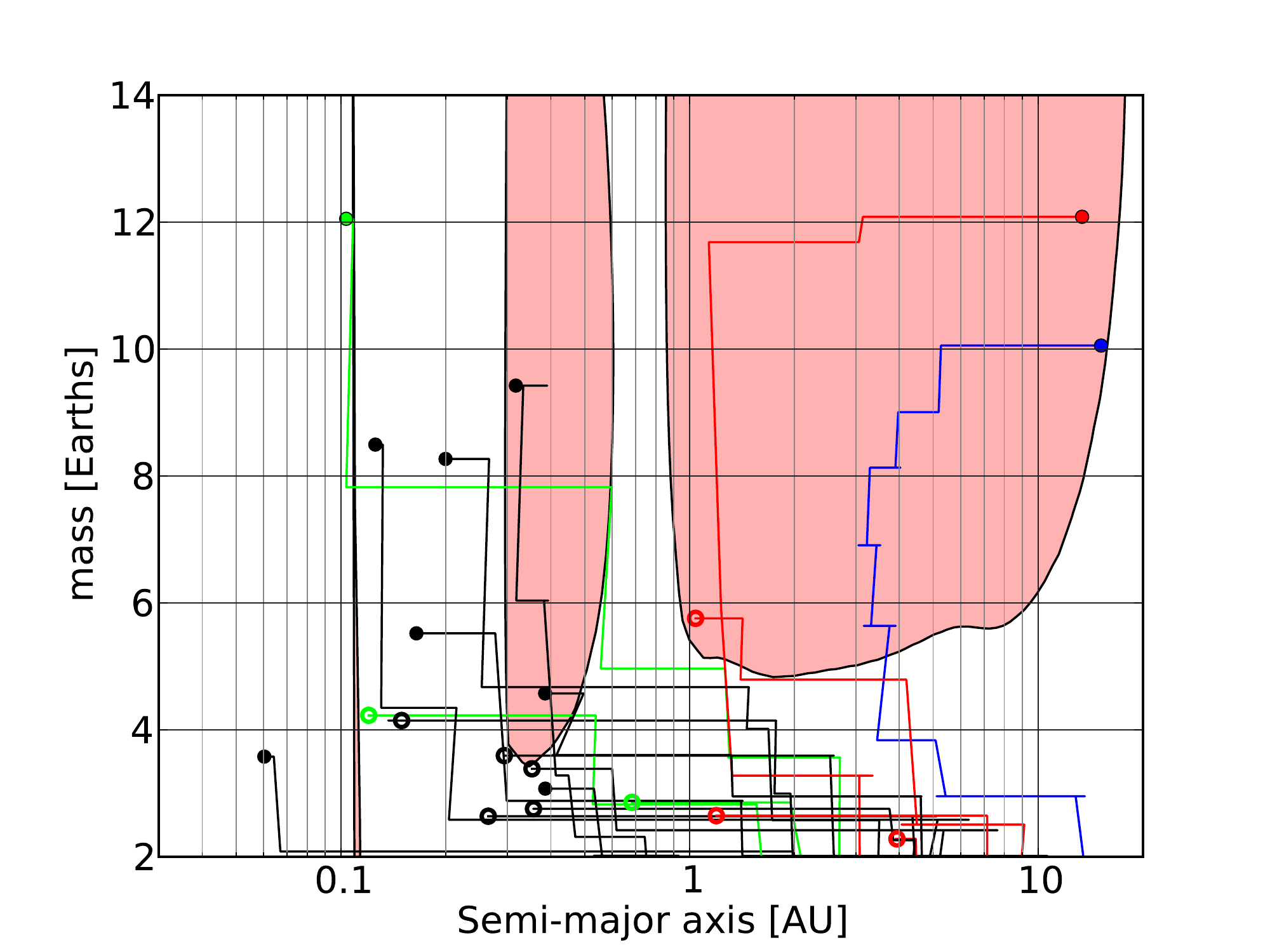}
 \caption{Mass vs. semimajor axis evolution of the embryos in a simulation that produced two giant planet cores.  Formatted as in Fig.~\ref{fig:GPC_case}.}\label{fig:twin_GPC}
\end{figure}

\reffig{fig:twin_GPC} shows a high-metallicity simulation that formed two giant planet cores.  In this case two embryos crossed into the outward-migration zone. They did not migrate monotonically outward, as repeated encounters with inward-migrating embryos imparted small inward kicks.  The two cores settled near the zero-torque zone locked in 6:5 resonance.  Note that the cores are actually shifted slightly interior to the zero-torque location because of resonant eccentricity excitation~\citep{cossou2013convergence}.  The inner core is more massive than the outer one even though it crossed into the outward-migration region later.  This is simply because it underwent multiple large collisions closer-in.  By the end of the simulation the inner part of the system has formed a rich system of seven hot super-Earths/mini-Neptunes with masses of 3.5 to $12 \mearth$.  

\begin{figure}[htb]
\centering
\subfloat[Low metallicity (Total starting mass in embryos $M_\text{tot}=21\unit{M_\oplus}$)]{\label{fig:low_metallicity}\includegraphics[width=0.45\textwidth]{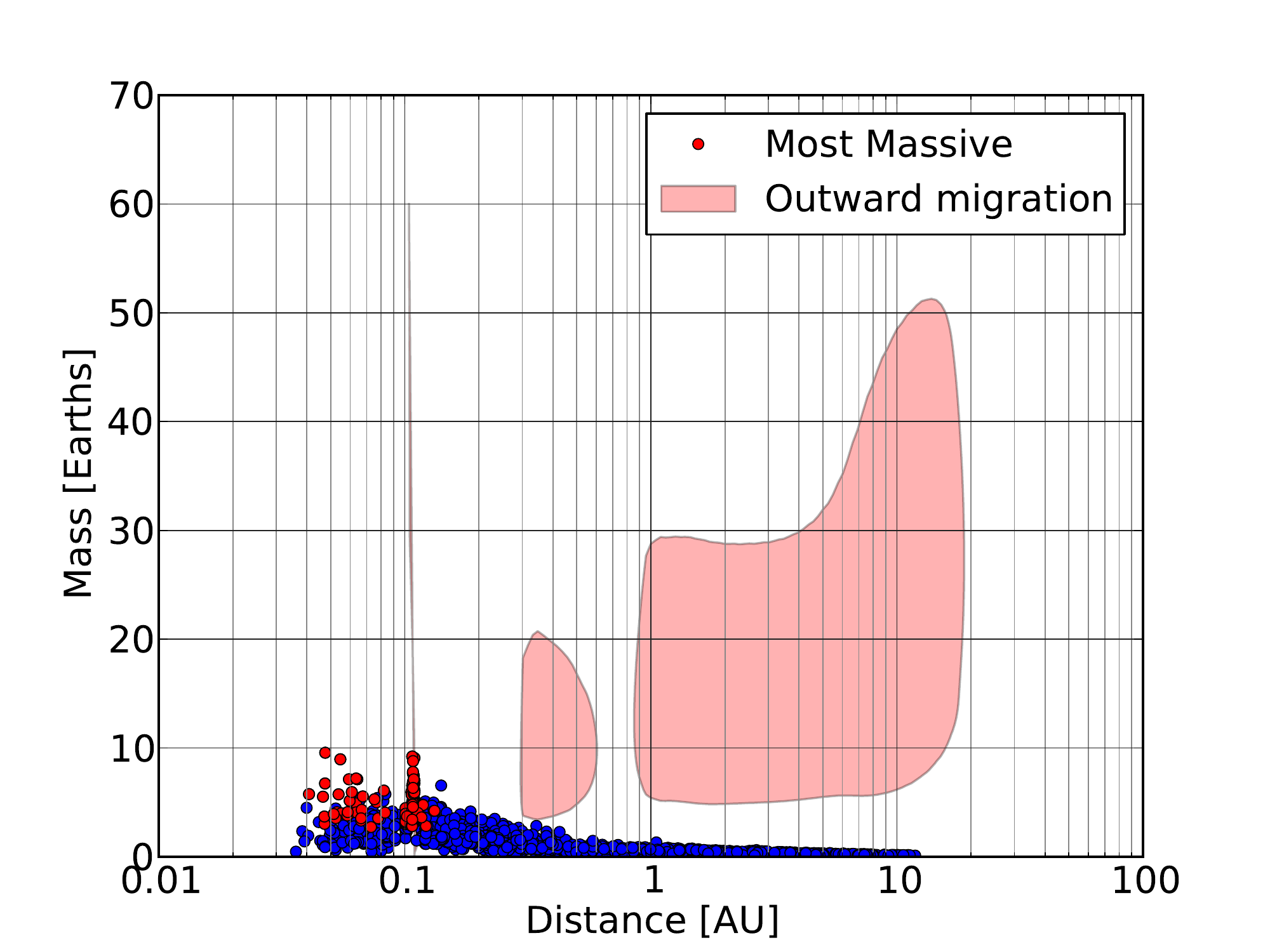}}\hfill
\subfloat[High metallicity ($M_\text{tot}=84\unit{M_\oplus}$)]{\label{fig:high_metallicity}\includegraphics[width=0.45\textwidth]{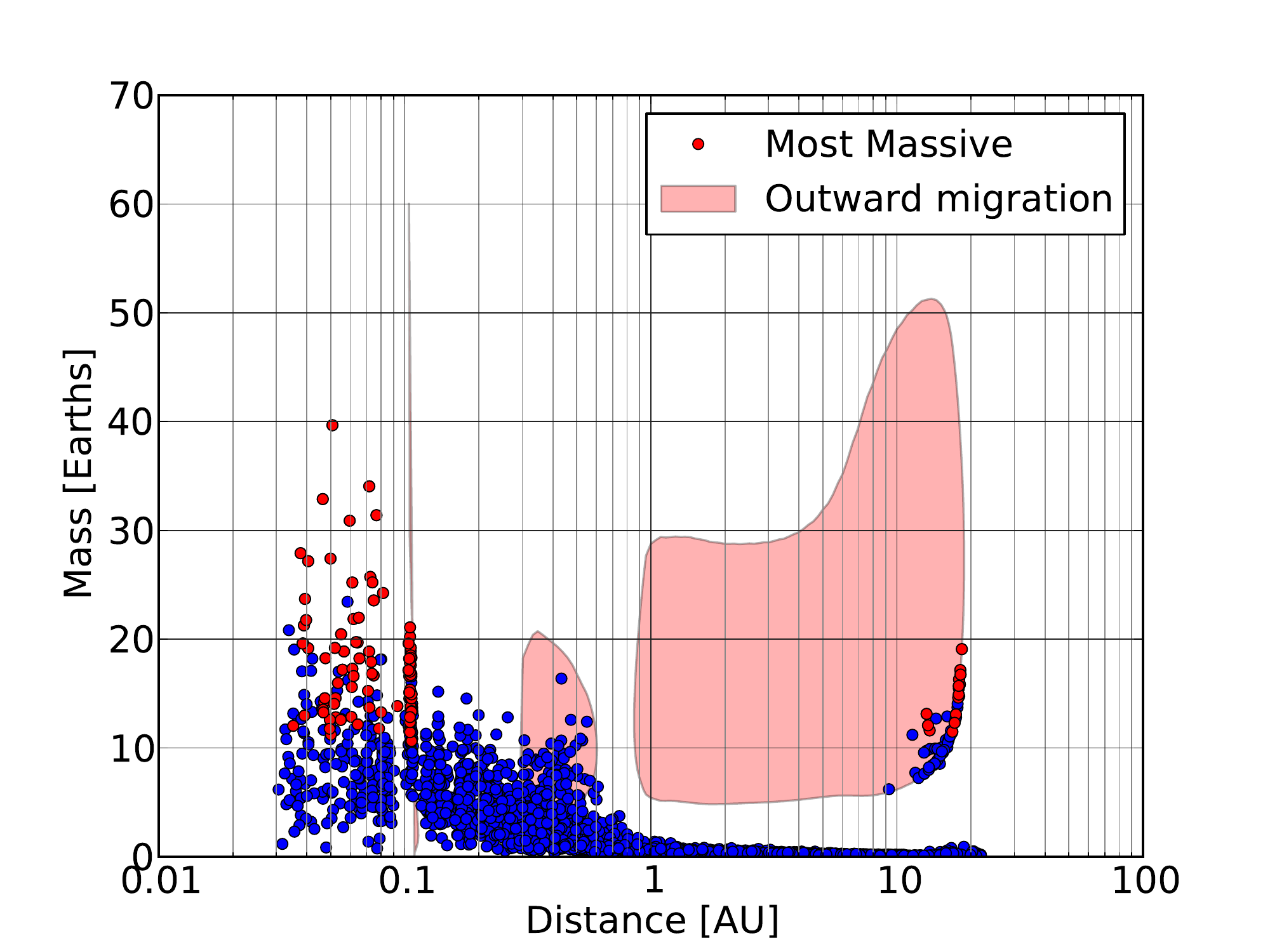}}
\caption{Final configuration of simulations with half (top panel) and twice (bottom panel) the fiducial metallicity, meaning with half or twice as much initial mass in planetary embryos. \modif{The disk is static such that this migration map applies throughout the simulation.} The outcomes of the fiducial simulations were shown in \reffig{fig:fiducial_static_stat}}\label{fig:metal-influence}
\end{figure}

Figure~\ref{fig:metal-influence} shows the final configuration of the simulations with half and twice the metallicity of our fiducial case.  Giant planet cores form far more readily in the high-metallicity systems. Collisions are more frequent given the larger mass reservoir, so it is more common for an embryo to jump into the outward migration regime.  This is especially important since the disk lifetime is limited, and embryos themselves may take longer to form in low-metallicity environments.  

\begin{table*}
\centering
\begin{tabular}{|c|c|c|c|}
\hline 
 & Low  & Fiducial & High \\
Constraint & Metallicity & & Metallicity\\
\hline
$a>1.5$ AU ; $m>6\mearth$ & 0\% & 12\% (13 candidates) & 73\% (118 candidates) \\ 
\hline 
$a>1.5$ AU ; $m>4\mearth$ & 1\% (1 candidate) & 65\% (90 candidates) & 100\% (384 candidates) \\ 
\hline  
Surviving at 3 Myr & 0 & 5 (all single cores) & 60 (14 with 2 cores) \\
\hline
\end{tabular} 
\caption{The fraction of simulations that formed giant planet core candidates.  The first row shows the number of cores that entered the outward-migration zone (defined at left).  The second row shows the number of cores that came within 33\% in mass of the outward-migration zone.  The third row shows the number of cores that actually survived in that zone at the end of the simulations. }\label{tab:frequency_GPC}
\end{table*}

\reftab{tab:frequency_GPC} lists the frequency of forming a giant planet core for the three different stellar metallicities.  This includes both the actual cores that were produced in the simulations and the fraction of simulations in which a core passed within 33\% in mass of the outward migration zone.  We consider the simulations to be the lower limit on the efficiency of core formation.  The number of cores that approach the outward-migration zone should provide a rough upper limit to the efficiency of core formation, allowing for additional unaccounted sources of accretion.  

The high-metallicity simulations were the only ones to produce systems with multiple giant planet cores.  The broad eccentricity distribution of the observed giant exoplanets~\citep{butler06,udry07} can be naturally explained if many systems of giant planets undergo dynamical instabilities~\citep{chatterjee08,juric08,raymond10}.  Hot Jupiters with orbits that are misaligned with the stellar equator~\citep{triaud10,winn10} may be explained by extreme instabilities in which one planet's eccentricity becomes so high that it undergoes tidal interactions with the host star~\citep{nagasawa08,beauge12}.  Clearly, systems with multiple gas giants are needed to explain these occurrences.  The observed planet-metallicity correlation is certainly biased toward close-in gas giants.  It has also been shown that is giant planets orbiting high-metallicity stars that show signs of planet-planet interactions and thus must form multiple systems~\citep{dawson13}.  Our model qualitatively explains how this might occur by preferentially producing systems with multiple giant planet cores around high-metallicity stars.  

Apart from the frequency of giant planet cores, the low- and high-metallicity simulations evolved in a similar way.  In both sets, hot super-Earths were ubiquitous.  The broad characteristics of the hot super-Earths were also similar although the planets' masses were larger in the high-metallicity systems.  Apart from a few outliers, the typical scaling was slightly weaker than linear with the metallicity: the median planet mass within 0.2 AU was 2.4/4.6/7.9~$\mearth$ for the low/standard/high metallicity simulations.  

To summarize, our model qualitatively reproduces the observed trends.  The frequency of giant planet core formation is a strong function of the metallicity.  However, the formation of hot super-Earths is not sensitive to metallicity.  

\section{Simulations with dissipating disks}\label{sec:dissip}

Real protoplanetary disks are not static.  Rather, they spread out under the effect of internal viscosity~\citep{lyndenbell1974evolution,pringle1981accretion,lin1990formation}.  Most of the disk falls onto the growing star and produces an observable accretion signature~\citep{muzerolle03,muzerolle05}.  In the late phases of evolution, disks are photo-evaporated both by their central stars~\citep{hollenach1994photoevaporation,alexander06,owen2010radiation} and by any nearby OB stars~\citep{adams2000theoretical,throop2008tail}.  The late phases of evolution are thought to proceed extremely quickly, with a final dissipation that lasts only $\sim 10^5$ years\\\citep{wolk1996search,simon1995disk,chiang07,currie09,owen2010radiation,alexander13}.

We now take disk dissipation into account in a simplified way.  As described above, we performed simulations for 100 Myr in which the disk surface density decreased uniformly.  The surface density first decreased exponentially with a timescale $t_1 = $2 Myr.  This was designed to roughly mimic a slowly-evolving viscous disk.  The next phase started at 3 Myr and involved rapid exponential dissipation with timescale $t_2 = $50,000 years. This was intended to mimic the late, fast phases of dissipation.  \modif{During the disk dissipation, the disk migration map was re-calculated whenever the surface density changed by more than 10\%.  This corresponded to every 200,000 years at the start of the simulation and every 5,000 years during rapid dissipation. Computing the disk full thermal profile is computationally expensive.  By restricting the frequency of these calculations we sped up the code with no noticeable loss in accuracy.  }

\begin{figure*}[htb]
\centering
\subfloat[$T=0\unit{Myr}$]{\includegraphics[width=0.45\textwidth]{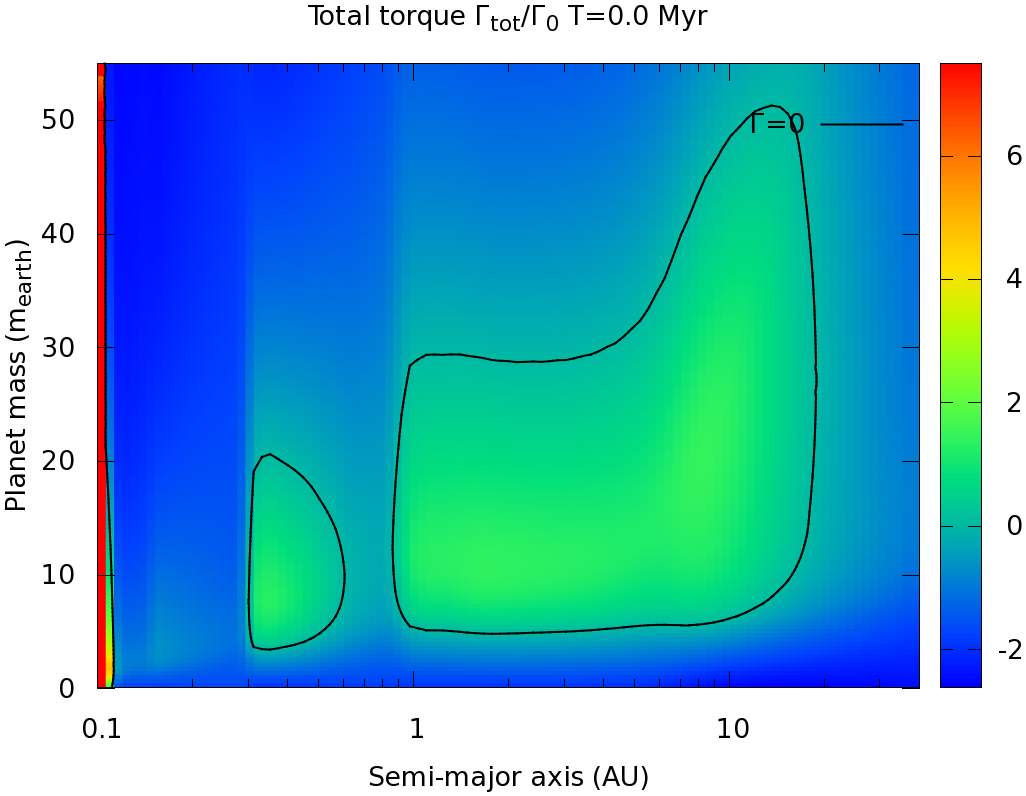}}\hfill
\subfloat[$T=2\unit{Myr}$]{\includegraphics[width=0.45\textwidth]{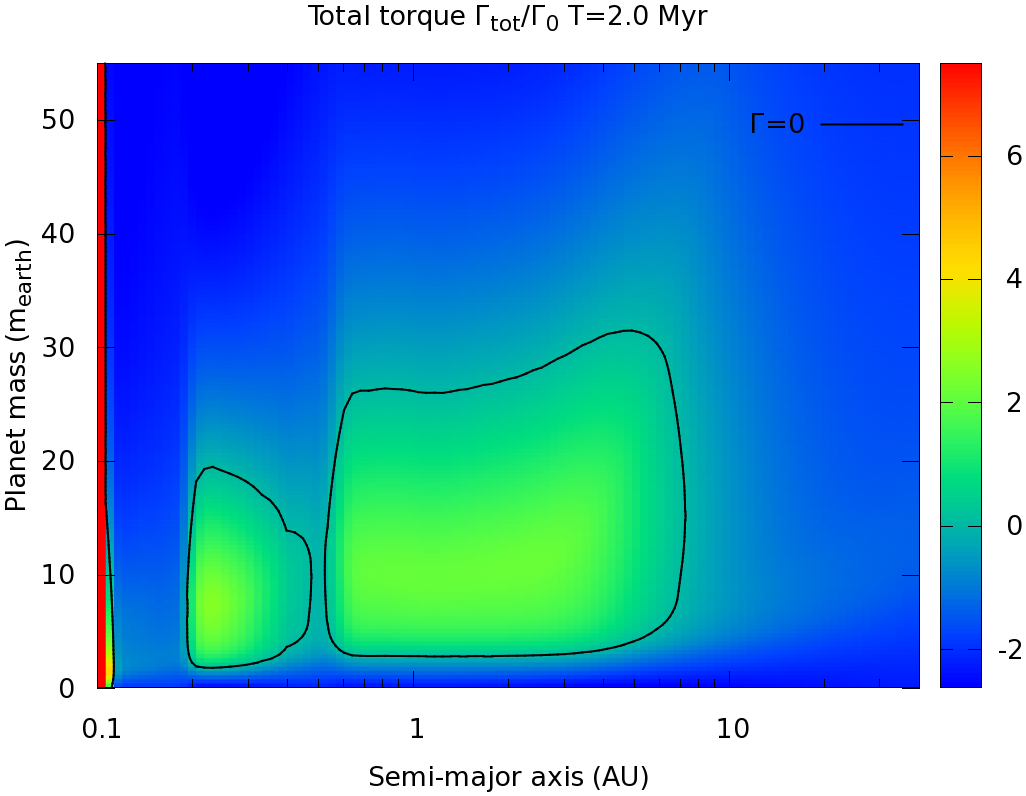}}

\subfloat[$T=3\unit{Myr}$]{\includegraphics[width=0.45\textwidth]{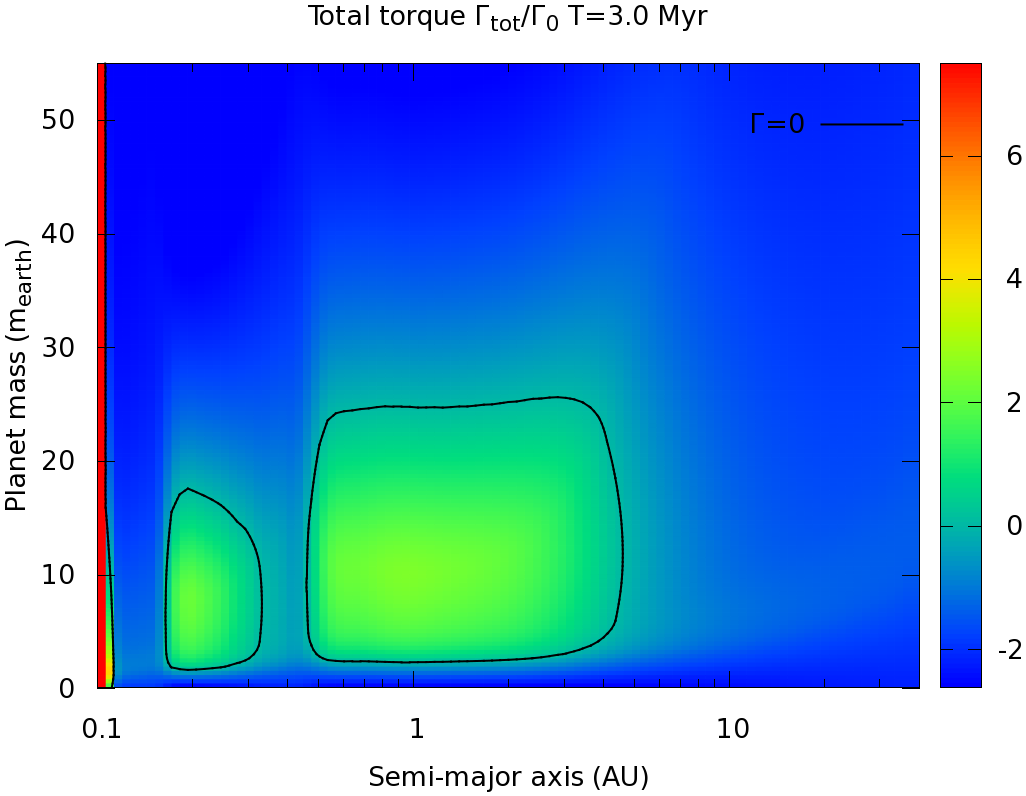}}\hfill
\subfloat[$T=3.4\unit{Myr}$]{\includegraphics[width=0.45\textwidth]{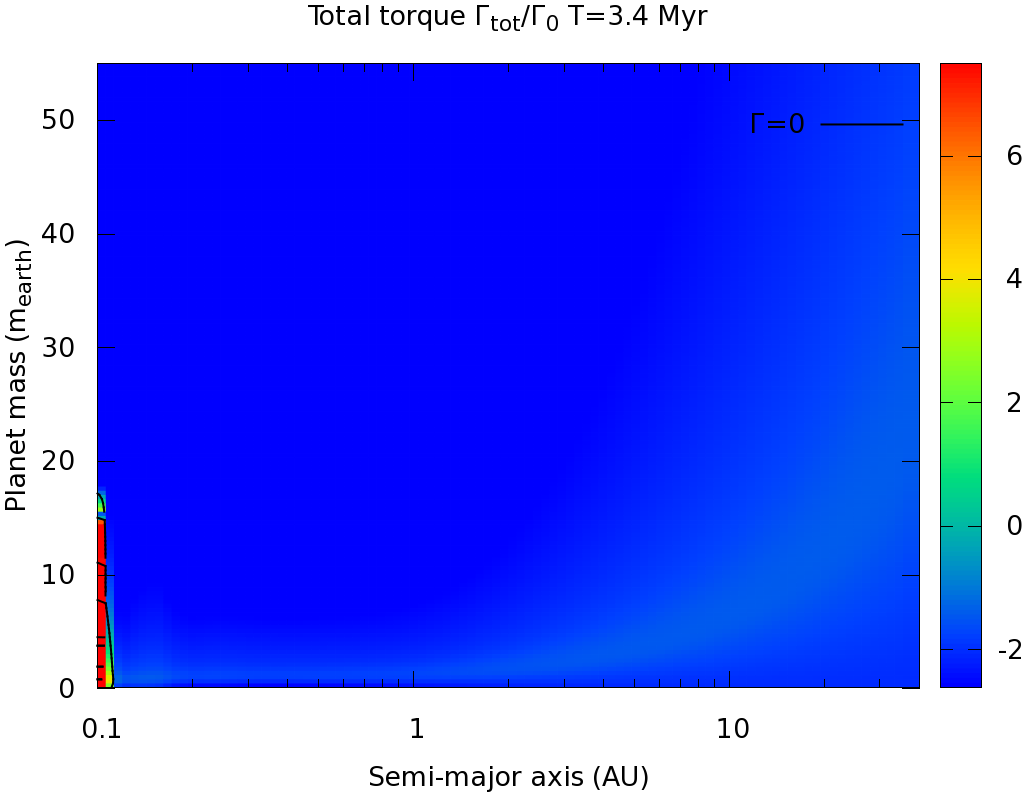}}
\caption{Evolution of Type I migration with dissipation of the disk. }\label{fig:dissipation_torque_evolution}
\end{figure*}

\reffig{fig:dissipation_torque_evolution} shows snapshots of the disk migration map during its evolution.  As the disk dissipates, the main zone of outward migration shifts inward and to lower masses.  Once the density drops below a critical threshold, the zone of outward migration disappears entirely~\citep[see also][]{bitsch2014stellar}.  

\begin{figure}[htb]
 \centering
 \includegraphics[width=0.85\linewidth]{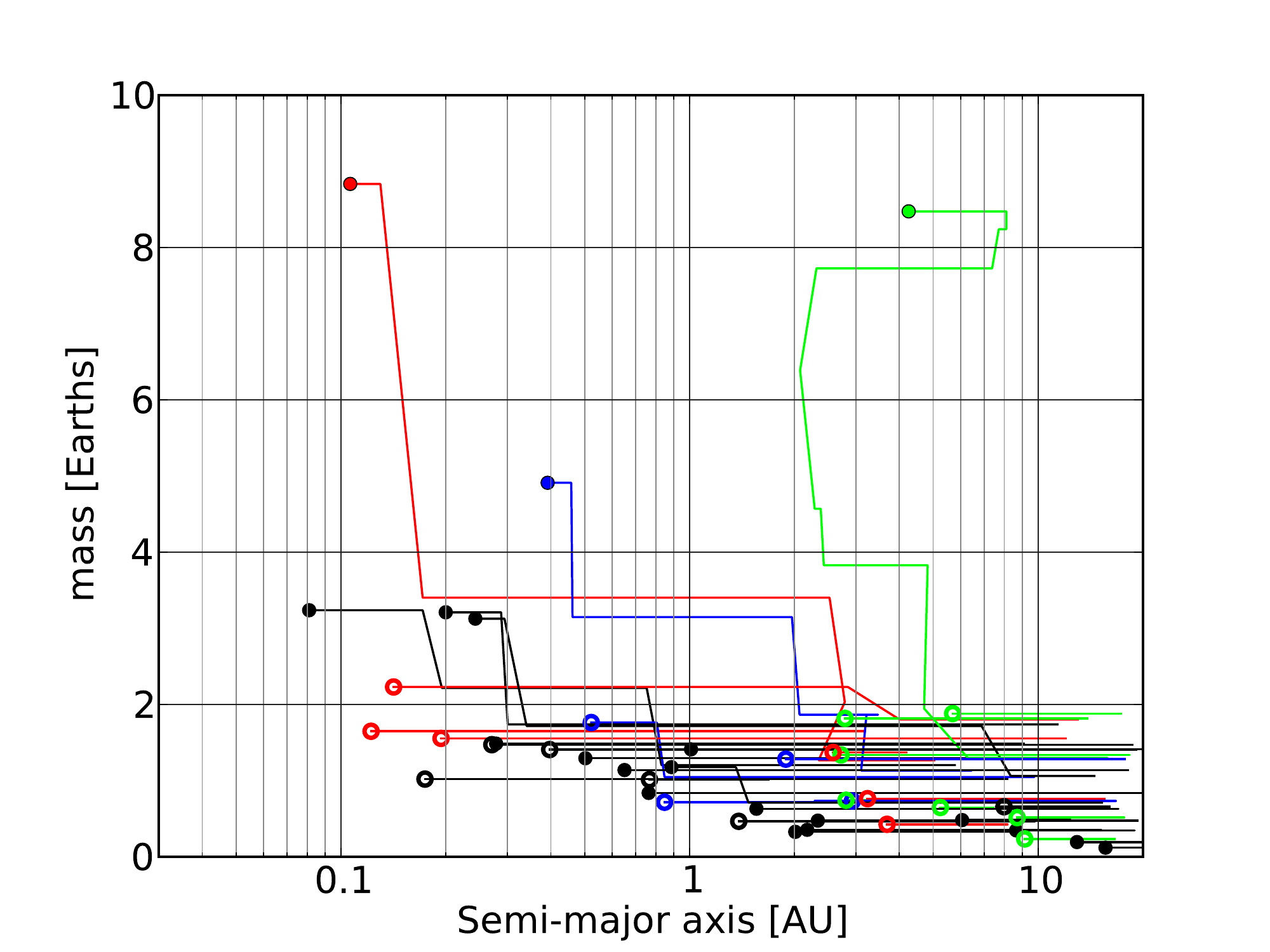}
 \caption{Mass vs. semimajor axis evolution of the embryos in a simulation that produced one giant planet core that later migrated inward as the disk dissipated.  Formatted as in Fig.~\ref{fig:GPC_case}.}\label{fig:dissip_simulation}
\end{figure}

Figure~\ref{fig:dissip_simulation} shows the evolution of a simulation in a dissipating disk. The early phases of evolution are similar to simulations in static disks.  Embryos migrate inward, collide and grow.  In this simulation a giant planet core grew and migrated to a zero-torque zone, but that zone itself migrated inward.  The core finished the simulation closer in than in the simulations with static disks. The core migrated with the zero torque zone until the zero torque zone disappeared. \modif{This is similar to the simulations of \cite{lyra10}, in which cores migrated inward in zero torque zones until the timescale of orbital migration exceeded the timescale for the evolution of the zero torque zone. The difference between the models is that the zero torque zone in our model disappears at late times whereas \cite{lyra10}'s does not.}

As the disk dissipated, the eccentricity and inclination damping felt by the super-Earths decrease, and the late stages of dissipation triggered an instability among the orbits of the close-in super-Earths~\citep[see also][]{iwasaki01,kominami04}.  This led to a late stage of embryo-embryo collisions with no gas present to damp embryos' random velocities.  This essentially marks a transition to the in-situ accretion regime, which has been shown to reproduce a number of characteristics of the observed hot super-Earths, albeit with {\it ad hoc} initial conditions~\citep{hansen13}.  At the end of the simulation the system still contains 6 planets more massive than $1 \mearth$ within 0.5 AU.  However, because of this late instability the planets are not in resonance as they were in the static disk simulations.  

\begin{figure}[htb]
\centering
\includegraphics[width=0.9\linewidth]{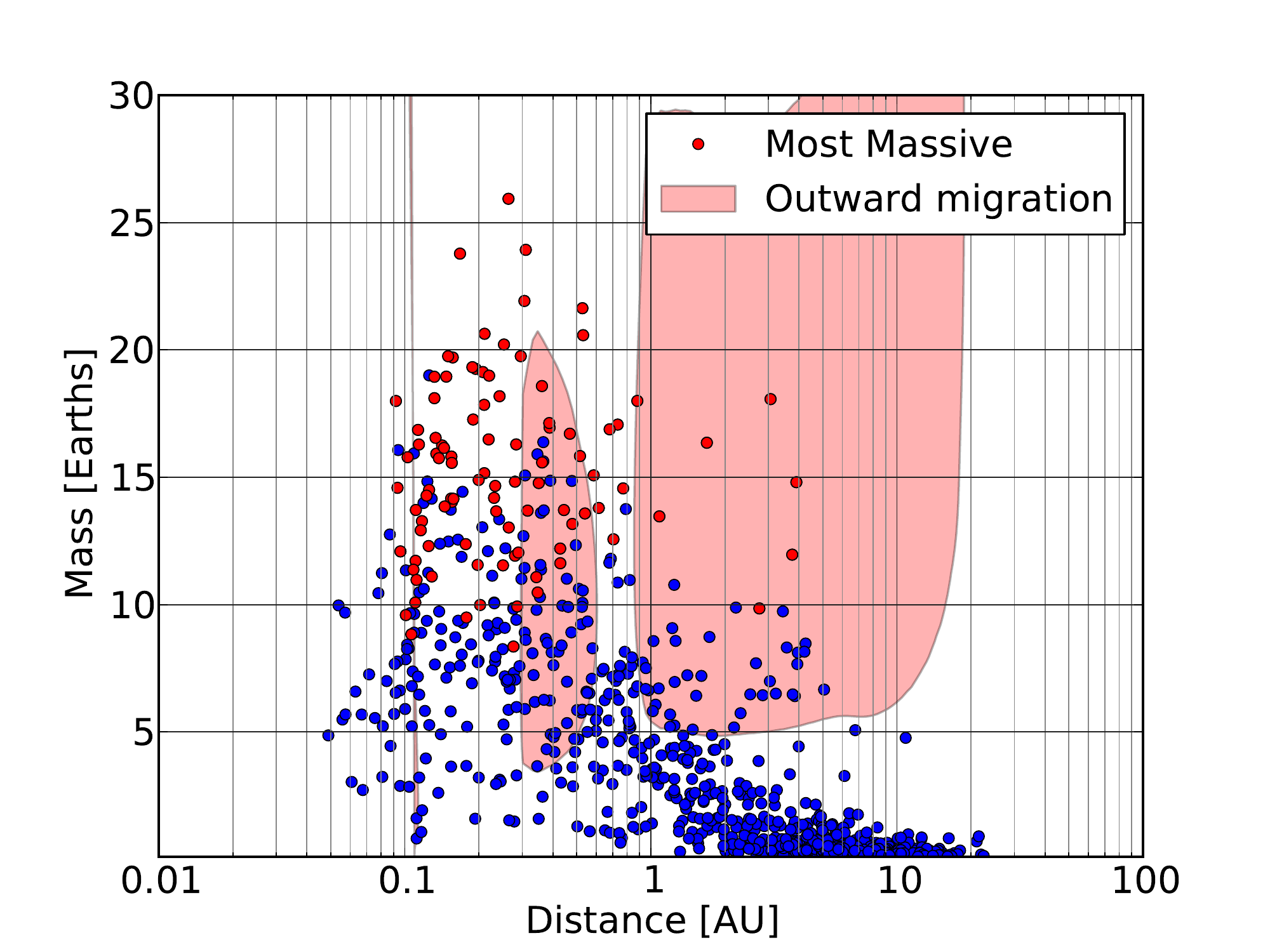}
\caption{Distribution of surviving planets in 100 simulations in dissipating disks.  Each simulation was run for 100 Myr. \modif{The migration map represents the one applied at the beginning of the simulation.} }\label{fig:slow_dissipation}
\end{figure}

Figure~\ref{fig:slow_dissipation} shows the distribution of planets that formed in 100 simulations with dissipating disks.  There are two main differences when compared with the planets that formed in static disks (Fig~\ref{fig:fiducial_static_stat}).  First, the planets in the inner part of the disk are less numerous and extend to higher masses.  This is simply because the instabilities triggered by the disk dissipation reduce the number of planets while increasing their masses.  Second, there are far more high-mass planets at orbital radii larger than 0.5-1 AU.  These are discussed in detail below.  

\subsection{Hot super-Earths}\label{sec:dissip_HSE}

We now compare the hot super-Earths that formed in our simulations within dissipating disks with a carefully-defined observed sample (see Section~\ref{sec:static_comp_obs}).  

\begin{figure*}[htb]
\centering
\subfloat[Orbital excitation]{\includegraphics[width=0.33\textwidth]{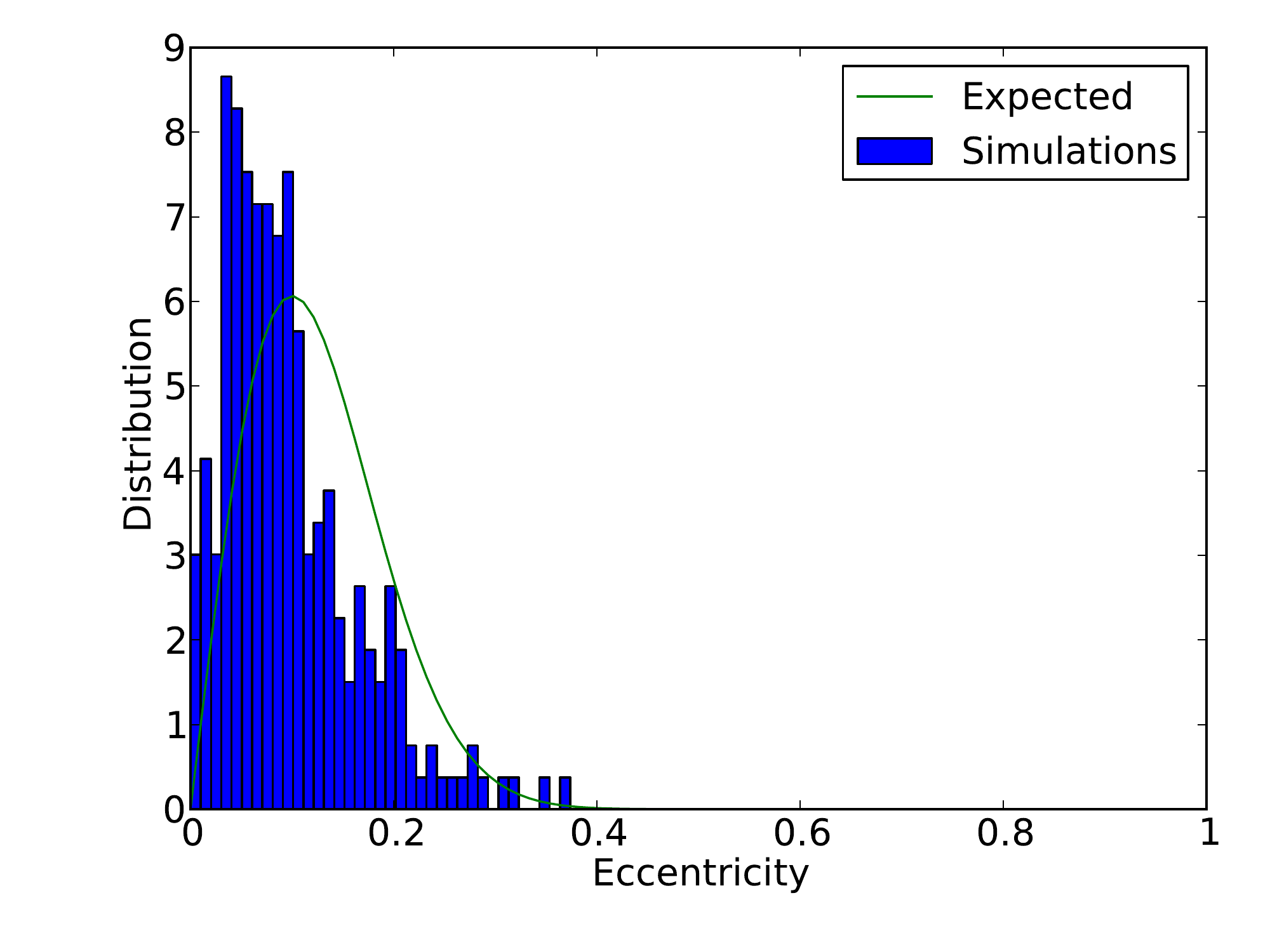}}
\subfloat[Coplanarity]{\includegraphics[width=0.33\textwidth]{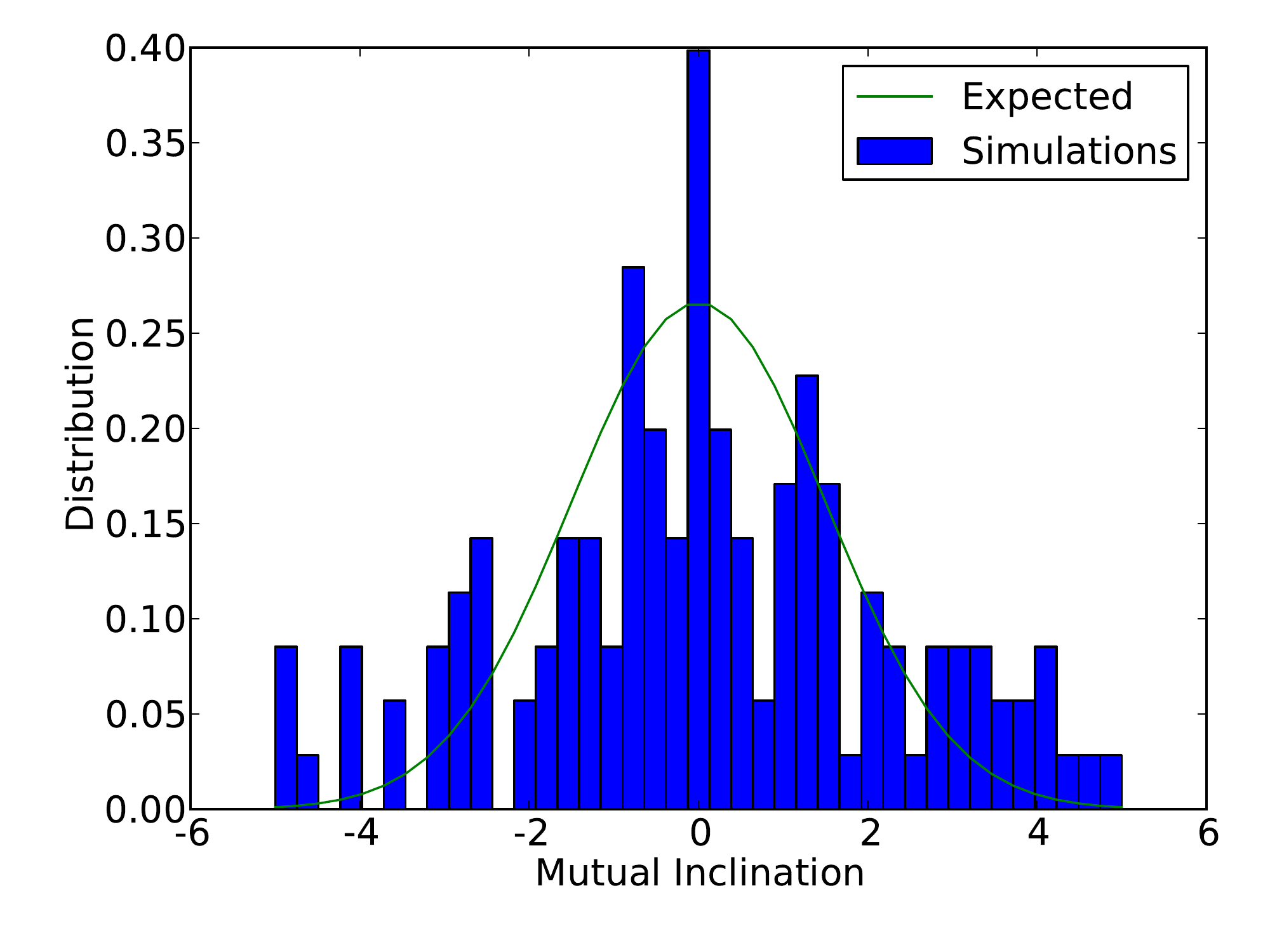}}
\subfloat[Mean motion resonances]{\includegraphics[width=0.33\textwidth]{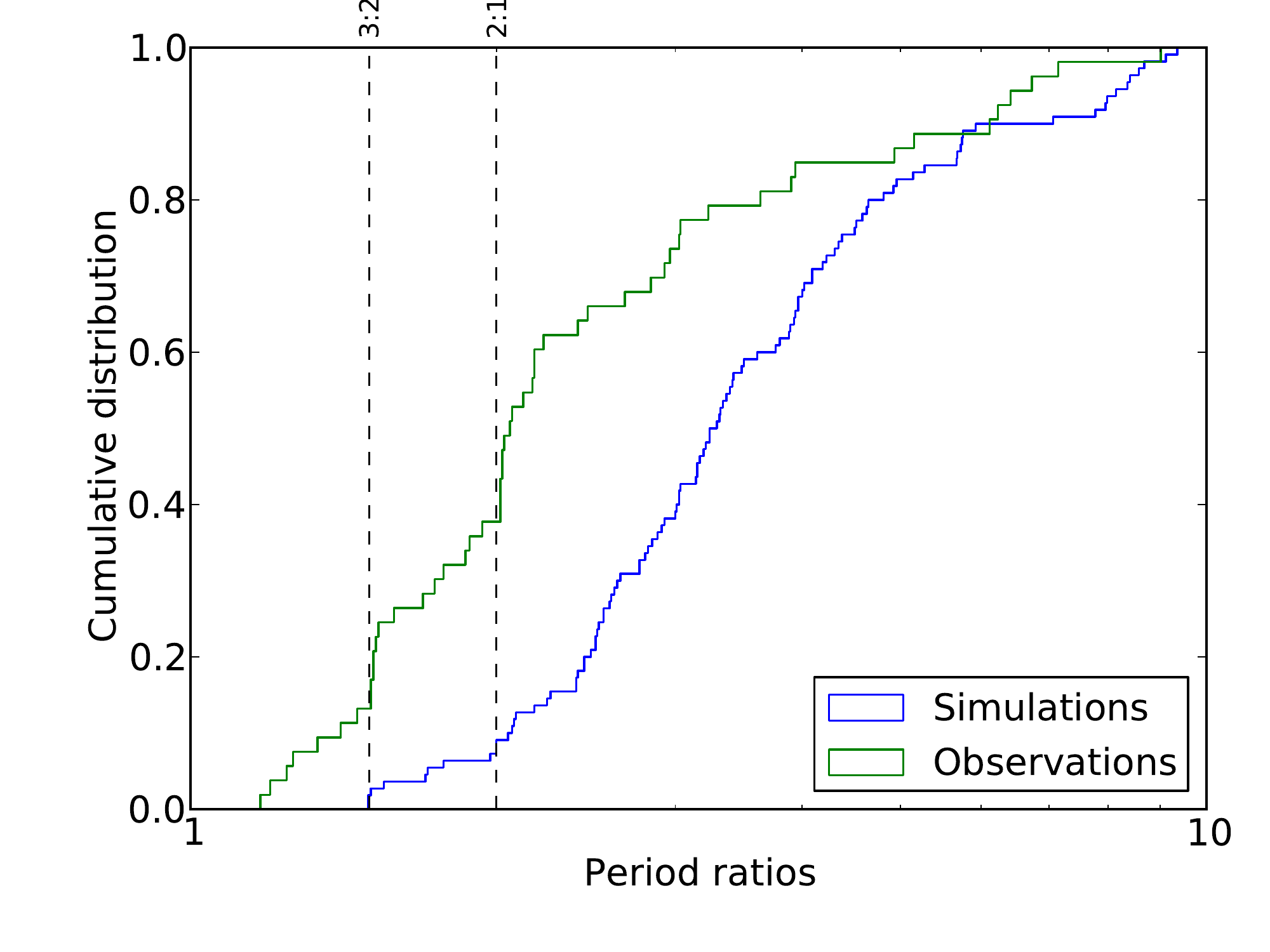}}
\caption{Comparison between the planets that formed in simulations with dissipating disks and observed extra-solar planets.  Both samples are limited to the range $R>1.5R_\oplus$ \modif{($3.3\unit{M_\oplus}$ assuming earth density)} and $p<200\unit{days}$.}\label{fig:comp_dissip}
\end{figure*}

Figure~\ref{fig:comp_dissip} shows a statistical comparison between our simulations in dissipating disks and the observations.  The inclination and eccentricity distributions are in good agreement with statistical inferences of the Kepler candidates~\citep{moorhead2011distribution,tremaine12,fang2012architecture}.  While the systems that formed in static disks had eccentricities and inclinations far smaller than expected, the systems that formed in dissipating disks provide a good match on both counts.  This is not surprising since {\it in-situ} accretion simulations have found the same thing~\citep{hansen13}, and the late phases of the simulations in dissipating disks essentially involve the same mechanisms as those at work in the in-situ accretion model.  

However, our simulations produce systems that are too spread out.  The distribution of the period ratios of adjacent planets does not match the observations.  The simulations in static disks provided a good match to the observed period ratio distribution.  The first few Myr of the simulations in dissipating disks are essentially identical to those in static disks. \modif{Although we did not explicitly simulate this, we expect that if the hot super-Earths formed in the static disk simulations were allowed to evolve for an additional 100 Myr in a gas-free environment, they would become unstable and their final distribution would match the hot super-Earths formed in the dissipating disk simulations.}  This means that the unstable late phase in the evolution of these systems caused them to spread out radially.  This is an expected consequence of an instability in a system of protoplanets given that collisions decrease the number of planets but not generally their radial extent.  

These simulations therefore match one important set of observables (the eccentricity and inclination distributions) but fall short on another (period ratio distribution).  \modif{However, as we discuss in section 6.3, it is probably very simple to match the period ratio distribution by invoking slightly faster inward migration.  Faster migration produces more compact resonant chains of close-in planets.  Late instabilities triggered by the disk dispersal can spread the systems out and match the observed period ratio distribution~\citep{cossou2013thesis}.}

\subsection{Giant planet cores}\label{sec:dissip_GPC}

In static disks it is easy to define giant planet cores as embryos that enter the region of outward migration beyond $\sim 1.5$~AU.  However, in a dissipating disk the region of outward migration itself migrates and changes shape (\reffig{fig:dissipation_torque_evolution}).  The two most important aspects of this evolution are that 1) the critical mass for outward migration decreases and 2) the zero-torque radius moves inward.  A core may therefore end up at a much closer-in orbital radius in a dissipating disk.  

Indeed, \reffig{fig:slow_dissipation} shows an abundance of planets of several to $\sim 20 \mearth$ at orbital distances of a few AU.  Whether these objects can really represent giant planet cores depends on how and when they formed.  To accrete and retain a thick gaseous envelope a core must reach a significant size at an early-enough time.  Sufficient gas must remain in the disk for gas accretion to occur.  The core's orbit must also be large enough.  Gas accretion is more efficient at large orbital separation~\citep{ikoma01} and indeed, the bulk of known giant exoplanets are located at $\sim$0.5-5 AU~\citep{butler06,udry07,mayor11}.

We apply a simple physically-driven criterion to identify potential giant planet cores.  We require cores to have grown to at least $5 \mearth$ at an orbital distance of 1.5~AU or larger in the first 2 Myr of the simulation.  At this point the gas density has dropped to $110\unit{g/cm^2}$ at $1\unit{AU}$.  The $5\mearth$ threshold was chosen as an approximate lower limit for efficient gas accretion~\citep{hubickyj2005accretion}.  

The formation of giant planet cores is as- or more efficient in dissipating disks than in static disks. \modif{To compare between simulations in static and evolving disks, we applied the same mass criterion: potential giant planet cores must be larger than $4\mearth$ although we stress that this chosen value is somewhat arbitrary.}

In the 100 static disk simulations there are 23 embryos in 21 systems that satisfy our three criteria.  If we slightly weaken the mass constraint to $4 \mearth$ to allow for additional sources of accretion (gas and planetesimals; see Section~\ref{sec:static_GPC}) this increases to 72 candidate giant planet cores in 53 systems.  In the 100 simulations in dissipating disks there are 29 candidate giant planet cores in 27 systems.  With the $4 \mearth$ mass constraint there are 83 candidate giant planet cores in 58 systems. \modif{The $4 \mearth$ criterion is probably optimistic with regard with outward migration in static disks (see Section~\ref{sec:static_GPC}). However, choosing $6 \mearth$ or $4 \mearth$ as mass criterion does not change the result: dissipating disks produce more giant planet cores than static disks.}

Unfortunately, we could not track the evolution of all of the candidate giant planet cores.  Many candidates would not become cores. Indeed, only 5 unambiguous cores survived among the 23 candidates in static disks. Most candidate cores were pushed inward -- interior to the outward-migration zone -- by inward-migrating embryos.  It is likely that additional effects would increase the fraction of candidates that become true cores.  Our simulations do not include gas or planetesimal accretion.  Even a modest increase in the mass of a candidate core would have two key effects.  It would strengthen its positive corotation torque.  And it would reduce the amplitude of its eccentricity excitation by inward-migrating embryos and thereby resist corotation torque damping~\citep{cossou2013convergence}.  

Among the surviving planets in the dissipating disk simulations (\reffig{fig:slow_dissipation}) there are 102 planets with $M \ge 5 \mearth$ and $a 
\ge 0.5$~AU.  29 of these satisfy our criteria as potential giant planet cores.  The others formed too slowly and are simply super-Earths at modest orbital separations. 

Formation of giant planet cores in dissipating disks is therefore another strong point of our model.  The frequency of the formation of candidate giant planet cores is at least 5\% and perhaps as high as $\sim 50\%$ \modif{(see \reftab{tab:frequency_GPC})}.  However, more detailed simulations including gas accretion onto growing cores and a feedback on the underlying gas disks are needed to quantify this.

\section{Discussion}\label{sec:discussion}

We now discuss our model's successes and challenges.  First, in Section 6.1 we summarize our model.  We discuss the requirements and assumptions inherent in our model.  In Section 6.2 we discuss the limitations of our simulations and prospects for improving them.  Finally, in Section 6.3 we discuss how well our model can reproduce specific observational constraints.   

\subsection{Summary of our model}

\begin{figure}[htb]
\centering
\includegraphics[width=\linewidth]{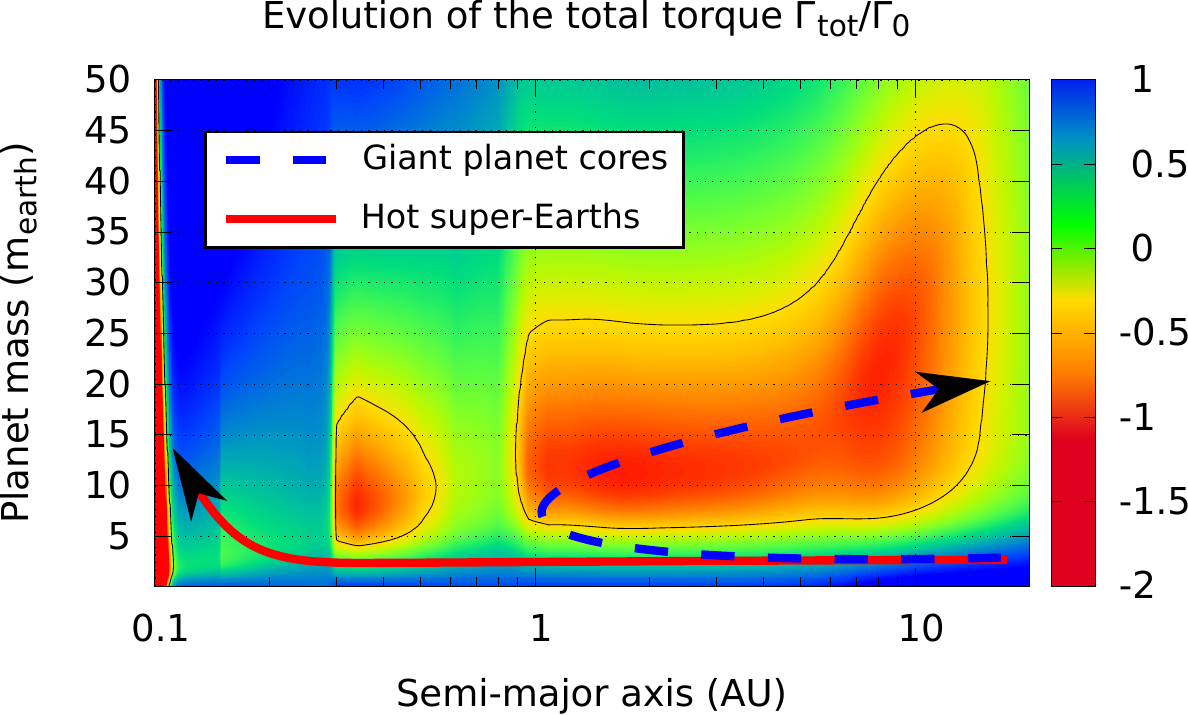}

\caption{Our formation scenario for hot super-Earths and giant planet cores. These scenarios are represented over the migration map of our reference disk, showed in \reffig{fig:migration_map}.}\label{fig:formation_paths}
\end{figure}

We propose that giant planet cores and hot super-Earths form from the same parent population (see \reffig{fig:formation_paths}).  Planetary embryos form at orbital radii of several to tens of AU and migrate inward.  If they grow slowly, embryos migrate all the way to the inner edge of the disk where they undergo collisions and pile up in resonant chains.  Embryos that become massive enough migrate outward to zero-torque radii and become giant planet cores.  As the disk dissipates, cores migrate inward with the zero-torque radii until they disappear.  The inner resonant chain of super-Earths is destabilized as the disk damping is removed.  There is a phase of collisions akin to the last phases of in-situ accretion.  The surviving super-Earths are no longer systematically found in resonances.  

Our model is appealing in its simplicity.  The same disk can form multiple types of planets depending on the simple competition between migration and growth.  It naturally explains how disk with a simple power-law surface density profile can end up with a big pile up of mass in the inner parts of the system.  By destabilizing resonant chains of migrated embryos during the dissipation of the disk our model retrieves the attractive aspects of the in-situ accretion model~\citep{hansen13} without invoking {\em ad hoc} or unphysical initial conditions.  

Our model requires that embryos form quickly at orbital distances of several AU.  While the formation of these objects is not fully understood, such cores must indeed form in order to explain the known gas giants and ice giants.  Our model also requires the existence of a zone in which embryos migrate outward.  Type I migration maps are sensitive to a range of disk parameters; a non-exhaustive list includes the total disk mass, viscosity, viscosity profile (alpha vs constant), the opacity table, the accretion rate and the planetary eccentricity~\citep{kretke12,bitsch13,bitsch2014stellar,cossou2013thesis}.  Nonetheless, a wide range of tested disk models do indeed contain zones of outward migration.  

\subsection{Limitations of the simulations}

We consider this study to be a proof of concept.  We have demonstrated that our model is valid and interesting but our simulations only covered a narrow range of parameter space and were missing several important effects.  We now discuss these limitations.  

Our simulations only consider a single, somewhat arbitrary disk profile.  Type I migration maps were produced for a vast range of disk parameters during C. Cossou's thesis~\citep[see also ][]{kretke12,bitsch13,bitsch2014stellar}.  Although there was a lot of diversity, a broad range of disks had similar structure.  In particular, the important outer lobe of outward migration was at a similar orbital radius and mass for a range of disk parameters.  The disk detailed in Section 3.1 has structure that is characteristic of these disks.  Testing additional disk profiles would nonetheless help constrain the global validity of our model.  

We also only considered a single, oversimplified mode for dissipation of the disk.  New models show that the dissipation of the disk is complex, time-varying and inside-out or, in some cases, outside-in~\citep[see review by][]{alexander13}.  The most important aspect of the disk dissipation is to trigger instabilities in the resonant chains formed toward the inner edge, so we do not expect this to strongly affect our results.  Nonetheless, it would be interesting to include a more realistic treatment of disk dispersal in this type of simulation.  

There are several key effects missing from our simulations.  Probably the most important is that we do not include gas accretion onto the growing embryos.  Gas accretion depends on the planet's mass and density and the disk local thermodynamic properties~\citep[e.g.][]{ikoma01,hubickyj2005accretion}.  Even a small increase in the embryo mass during its inward migration could have an important consequence, by possibly pushing it into the outward-migration region.  In addition, if any candidate giant planet cores undergo rapid gas accretion and become gas giants, they should carve annular gaps in the disk and transition to type II migration~\citep{lin96,ward97}.  This in turn affects the dynamics of the entire system~\citep[e.g.][]{thommes08,hellary2012global}.  

Likewise, we did not include planetesimals, pebbles or dust in the simulations.  A large reservoir of small bodies should have both dynamical and accretional consequences at early times.  Small bodies provide both efficient damping of random velocities~\citep{chambers06} and in some cases can be efficiently accreted~\citep{rafikov04,levison10,lambrechts12}.  

We also did not include collisional fragmentation.  A significant fraction of impacts between embryos should not lead to perfect merging~\citep{agnor04,asphaug06}.  Rather, there exists a wide range of possible outcomes~\citep{leinhardt12,genda12}.  Although this does not appear to strongly affect the outcome of late-stage accretion~\citep{kokubo10} it may affect how fast planets grow~\citep{chambers13}, a critical parameter for giant planet core formation. \modif{It remains unclear whether giant impacts have a negative effect on core growth due to collisional erosion or a positive one by promoting rapid gas accretion \citep{broeg2012giant}.}

Finally, our simulations did not include stochastic perturbations from turbulent fluctuations in the disk~\citep{laughlin2004type,ogihara2007accretion}.  Turbulence affects the stability of mean motion resonances~\citep{adams08,pierens13} and may play a role in shaping the period ratio distribution~\citep{rein12}. However, the huge number of planetary embryos ($\sim 100$) and perturbations that comes with it, act in a similar way and break resonances through planet-planet perturbations.

\subsection{Matching observations}

We now put our model in perspective by confronting it with several aspects of the observed extra-solar planet population.  We discuss both the specific distributions of certain planet classes and larger-scale issues.

\begin{itemize}

\item {\bf The inferred eccentricity and mutual inclination distributions of Kepler candidate super-Earths}~\citep{moorhead2011distribution,tremaine12,fang2012architecture}.  The planetary systems that formed in simulations in static disks were too dynamically cold, with very small eccentricities and inclinations.  However, simulations in dissipating disks produced systems with eccentricity and inclination distributions consistent with those expected from observations.  

\item{\bf The distribution of orbital period ratios of adjacent planets}~\citep{lissauer11b}.  Systems that formed in static disks survived in long resonant chains.  This is not what is observed~\citep[e.g.,][]{rein12}.  However, observational biases hide certain planets and the resulting distribution of expected detections is a good match to observations.  In contrast, the systems formed in dissipating disks provided a poor match to observations because the simulated systems were too spread out.  

In an earlier version of our code, embryos migrated too fast.  The type I migration timescale was half of its correct value.\footnote{We thank Aurelien Crida for finding this error.}  Many early sets of simulations were run with this code.  In those simulations we had the opposite problem of the one in the simulations presented here.  Simulations in static disks did not match the observations.  Rather, their orbital configurations were too compact.  Simulations in dissipating disks were spread out in comparison and did match the observations as well as the eccentricity and inclination distributions.  

For our model to match the period ratio distribution, embryos must simply migrate faster than in our simulations and pile up into more compact resonant chains (although the rate of eccentricity/inclination damping is also important for resonance capture).  This is a simple problem.  Faster migration could be achieved with a slightly more massive disk or one with a different surface density distribution.  \modif{One solution is for migration to occur slightly earlier in the disk evolution when it was presumably more massive.  After super-Earths are piled into a resonant chain, late destabilization would then reproduce the observed distributions. However, it is unclear whether a disk would remain massive for long enough for this to matter, as the phase during which observed disks have high accretion rates (and presumably high masses) only lasts a few hundred thousand years~\citep{hartmann98}}.  

\modif{Of course, the simplest solution would be if our simulations underestimated the type 1 migration rate for small planets.  Remarkably, this has recently shown to be the case!  A new study using 3-D simulations by \cite{lega2014migration} has in fact shown that low-mass planets migrate even faster than the rate used in our model.  }

\item{\bf The frequency of systems of hot super-Earths}.  The occurrence rate of super-Earths (or mini-Neptunes) around Sun-like stars is roughly 30-50\%~\citep{mayor11,howard12,dong13,fressin13}.  Every single one of our simulations produced a system of hot Super-Earths. This frequency is clearly too high.  There is no mechanism inherent in our simulations to stop this.  How can we reduce the efficiency of hot super-Earth formation?

One simple solution is for the gaseous disk to disperse very fast.  This should occur in strongly-irradiated stellar environments such as in very dense clusters.  We performed an additional set of simulations in which the disk dispersed exponentially with a single timescale of 500,000 years.   Figure~\ref{fig:fast_dissipation} shows the planets produced in those simulations.  Given the rapid disk dispersal, compact systems of hot super-Earths do not form at the disk inner edge simply because there is not enough time for embryos to migrate inward that far. Instead, the most massive planets end in the middle of the disk, around $1\unit{AU}$. Of course, what really matters in quenching the production of hot super-Earths is the time that the embryos can migrate.  Thus, if embryos simply form more slowly than assumed here then they would not be able to migrate as far and the outcome would be the same.  

\begin{figure}[htb]
\centering
\includegraphics[width=0.9\linewidth]{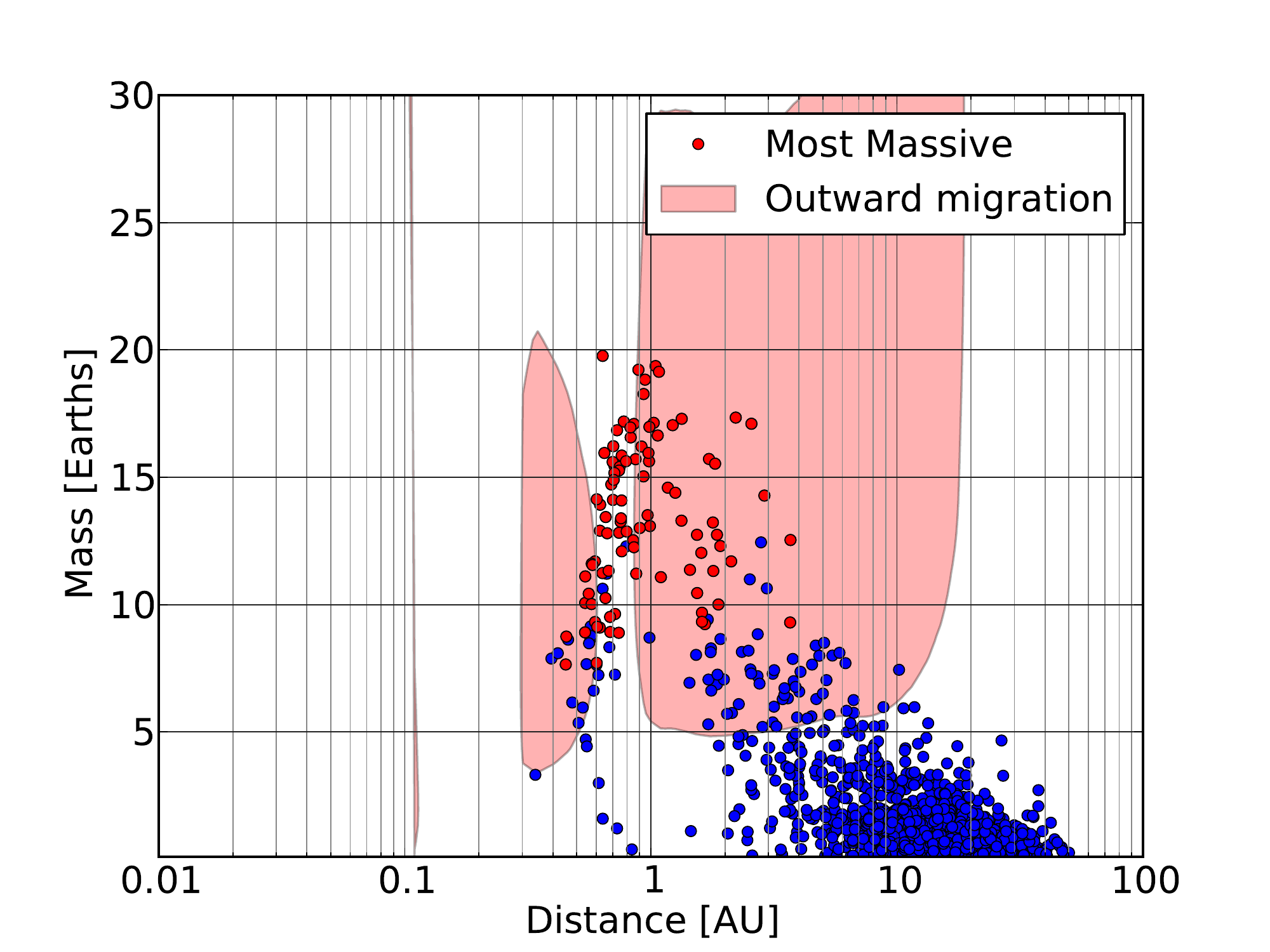}
\caption{Surviving planets in 100 simulations in which the gas disk dissipated fast, with an exponential decay with $\tau=500 000\unit{years}$.}\label{fig:fast_dissipation}
\end{figure}

Other mechanisms may exist to slow or stop super-Earth formation.  If a migration barrier existed at large orbital radius then this could simply stop inward-migrating super-Earths~\citep[e.g.][]{morby08}.  Such a barrier could be produced by a sharp opacity or density transition~\citep{masset06}.  More detailed disk models are required~\citep[e.g.,][]{bitsch2014stellar}.  Alternately, it is conceivable that gas giants at large orbital radius could hold up the inward-migrating embryos.  

\item{\bf The frequency of giant exoplanets}.  Gas giant planets exist with orbital periods less than 3000 days around $\sim$14\% of Sun-like stars~\citep{cumming08,mayor11}.  Most have orbital radii larger than 0.5-1 AU.  Our simulations do not include gas accretion onto growing embryos so we cannot make a direct comparison with observations.  

What we can constrain in our simulations is the efficiency of giant planet core formation.  In simulations with static disks, the number of actual cores that survive at or near the zero-torque zone provides a lower limit.  We obtained an upper limit on the efficiency of core formation by considering how many embryos passed either through or very close to the zone of outward migration.  This attempts to account for certain effects that are missing from our simulations, in particular, accretion of both planetesimals and gas.  If migrating embryos are able to grow faster then they can more easily enter the zone of outward migration and become candidate giant planet cores.  However, this issue is even more complex because the migration rates and planet-planet dynamics are strongly sensitive to the planet masses.  

Our fiducial simulations produced 5 surviving candidate giant planet cores in 100 simulations.  However, 13 embryos (in 12 simulations) entered the outward-migration zone and 90 embryos (in 65 simulations) came within 33\% in mass of the outward-migration zone \reftab{tab:frequency_GPC}.  This range in the potential efficiency of giant planet core formation is extremely wide, from 5\% to 65\%.  It is consistent with estimates of the frequency of giant exoplanets~\citep{cumming08,gould10,mayor11}.  Additional simulations that include gas accretion are needed to quantify this more firmly.  

\item{\bf The giant exoplanet-metallicity correlation}.  Several studies have demonstrated a strong correlation between the frequency of giant exoplanets and the stellar metallicity~\citep{gonzalez97,santos2004spectroscopic,fischer05}.  We tested the effect of metallicity in our model by performing two additional sets of simulations (in static disks), one with half and the other with twice the total initial mass in the embryo population.  We assumed that the total embryo mass was an adequate tracer of the dust-to-gas ratio and therefore the metallicity.  

Our simulations produced a much stronger than linear correlation between the stellar ``metallicity'' and the occurrence rate of giant planet cores (see \reftab{tab:frequency_GPC}).  This was true both for cores that survived at the end of the simulations and for those that passed through or near the outward-migration zone during their evolution.  Indeed, the low-, fiducial- and high-metallicity simulations produced embryos that entered the outward-migration zone in zero, 12\% and 73\% of simulations.  The high-metallicity simulations produced a large number of multiple-core (or multiple-candidate core) systems.  This is also consistent with the observational requirement that gas giants must form in multiple systems around high-metallicity stars, in order to explain their dynamical properties~\citep{dawson13}.

The frequency of hot super-Earths is not correlated with the stellar metallicity~\citep{ghezzi10,buchhave12,mann12}.  Our simulations also reproduce this observation.  Indeed, our simulations {\it always} form systems of hot super-Earths, regardless of the stellar metallicity.   

\end{itemize}

\begin{acknowledgements}
We thank the anonymous referee for helpful comments that improved the paper.  C. Cossou's thesis was generously funded by the Conseil Regional d'Aquitaine.  This research was also supported by the CNRS's PNP and EPOV programs and the Agence Nationale pour la Recherche (via grant ANR-13-BS05-0003-002; project MOJO). Computer time for this study was provided by the computing facilities MCIA (M\'{e}socentre de Calcul Intensif Aquitain) of the Universit\'{e} de Bordeaux and of the Universit\'{e} de Pau et des Pays de l'Adour.  We acknowledge many useful comments during C. Cossou's thesis defense from his committee of A. Crida, A. Morbidelli, C. Terquem and R. Nelson, and afterwards from B. Bitsch, A. Izidoro, F. Selsis and J. Szulagy.
\end{acknowledgements}

\bibliography{biblio,refs}

\end{document}